\documentclass[reprint, amsmath, amssymb, aip, jcp, onecolumn]{revtex4-2}

\usepackage{graphicx}
\usepackage{newtxtext}
\usepackage{newtxmath}
\usepackage{natbib}
\usepackage{hyperref}
\hypersetup{
    colorlinks = true,
    urlcolor   = blue,
    citecolor  = black,
}

\newcommand{\RomanNumeralCaps}[1]
\linenumbers

\usepackage[T1]{fontenc}
\usepackage[utf8]{inputenc}

\usepackage[version=3]{mhchem} 

\usepackage{multirow}
\usepackage{dcolumn}
\usepackage{bm}
\usepackage{subcaption}

\usepackage[normalem]{ulem}
\usepackage{makecell} 
\usepackage{booktabs}
\usepackage{xspace}

\usepackage[per-mode=reciprocal, range-units=single, range-phrase=--]{siunitx} 
\DeclareSIUnit[number-unit-product = {}]\sinumber{{\scriptstyle\#}}

\usepackage{cleveref}
\crefname{figure}{figure}{figures}
\Crefname{figure}{Figure}{Figures}
\crefname{table}{table}{tables}
\Crefname{table}{Table}{Tables}
\crefname{equation}{equation}{equations}
\Crefname{equation}{Equation}{Equations}

\usepackage[textsize=tiny,prependcaption,colorinlistoftodos,textwidth=1.5cm]{todonotes}   

\def \rb {\mathbf{r}}
\def \dr {\, \mathrm{d}\mathbf{r}}

\DeclareSIUnit\bar{bar}
\DeclareSIUnit\angstrom{\text{\AA}}

\def \vb {\mathbf{v}}
\def \d {\, \mathrm{d}}
\def \D {\mathrm{D}}

\def \res {\mathrm{res}}
\def \ext {\mathrm{ext}}
\def \jdiff {\mathbf{j}_i^\mathrm{diff}}

\def \diff {\mathrm{diff}}
\def \jdiff {\mathbf{j}_i^\mathrm{diff}}
\def \mw {M}
\def \kb {k_\mathrm{B}}

\def \divd {\nabla \!\boldsymbol{\cdot}\!}

\makeatletter
\newcommand{\thickhline}{%
    \noalign {\ifnum 0=`}\fi \hrule height 1pt
    \futurelet \reserved@a \@xhline
}
\newcolumntype{"}{@{\hskip\tabcolsep\vrule width 1pt\hskip\tabcolsep}}
\makeatother

\makeatletter
\providecommand\barcirc{\mathpalette\@barred\circ}\def\@barred#1#2{\ooalign{\hfil$#1-$\hfil\cr\hfil$#1#2$\hfil\cr}}
\newcommand\stst{^{\protect\barcirc}}
\makeatother

\begin{document}

\title{Mass Transfer Through Vapor-Liquid Interfaces From Hydrodynamic Density Functional Theory}

 \author{Benjamin Bursik}
\affiliation{Institute of Thermodynamics and Thermal Process Engineering, University of Stuttgart, Pfaffenwaldring 9, 70569 Stuttgart, Germany}
\author{Frederic Bender}
\affiliation{Institute of Thermodynamics and Thermal Process Engineering, University of Stuttgart, Pfaffenwaldring 9, 70569 Stuttgart, Germany}
\author{Rolf Stierle}
\affiliation{Institute of Thermodynamics and Thermal Process Engineering, University of Stuttgart, Pfaffenwaldring 9, 70569 Stuttgart, Germany}
\author{Gernot Bauer}
\affiliation{Institute of Thermodynamics and Thermal Process Engineering, University of Stuttgart, Pfaffenwaldring 9, 70569 Stuttgart, Germany}
\author{Joachim Gross}
\affiliation{Institute of Thermodynamics and Thermal Process Engineering, University of Stuttgart, Pfaffenwaldring 9, 70569 Stuttgart, Germany}
 
\begin{abstract}

  We assess the capabilities of hydrodynamic density functional theory (DFT) to predict mass transfer across vapor-liquid interfaces by studying the response of an initially equilibrated pure component vapor-liquid system to the localized insertion of a second component.
  Hydrodynamic DFT captures the effect of interfaces on the dynamics by modeling the chemical potential gradients of an inhomogeneous system based on classical DFT.
  Hydrodynamic DFT effectively connects molecular models with continuum fluid dynamics. Away from interfaces the framework  simplifies to the isothermal Navier-Stokes equations.
  We employ Maxwell-Stefan diffusion with a generalized driving force to model diffusive molecular transport in inhomogeneous systems. 
  For the considered Lennard--Jones truncated and shifted (LJTS) fluid, we utilize a non-local Helmholtz energy functional based on the perturbed truncated and shifted (PeTS) equation of state.
  The model provides noise-free partial densities and fluxes for the mass transfer near the interface, as well as profiles of these quantities across the interface.
  A comparison with non-equilibrium molecular dynamics simulations shows that hydrodynamic DFT accurately predicts mass transfer across the interface, including microscopic phenomena such as the temporary enrichment and repulsion of the light boiling component at the interface.
  Combining generalized entropy scaling with generalized Maxwell-Stefan diffusion allows for an accurate description of diffusive molecular transport in the system.
This approach can accurately predict phase behavior, equilibrium interfaces, and mass transfer across interfaces based on molecular interactions for mixtures of strongly dissimilar components.
Our results suggest that hydrodynamic DFT can accurately predict the dynamics of mixtures at vapor-liquid interfaces. 
This is important for modeling transport processes in fluid systems and porous media -- particularly for describing evaporation from pores, which requires accurate modeling of mass transfer across vapor-liquid interfaces.

\end{abstract}

  \maketitle

  \section{Introduction}
  
  Mass transfer across vapor-liquid interfaces is a fundamental phenomenon in natural and industrial processes. 
  In nature, it governs the exchange of light boiling components, such as carbon dioxide, oxygen, and methane, between the ocean and the atmosphere \citep{liss2014short}, as well as the growth of droplets during cloud formation \citep{zheng2018modeling}. 
  Furthermore, mass transfer across vapor-liquid interfaces plays a central role in the evaporation of liquids from porous media  \citep{ho1995mass,chaoyang1993twophase} and in distillation \citep{baur2005influence} or absorption/desorption processes \citep{dumont2003mass}.
  
  Several macroscopic mass transfer models, e.g.\ two-film theory, have been proposed. These models treat the interface as  infinitely thin and typically assume a negligible resistance to mass transfer across it \citep{wen2020fundamentals,doran2013chapter}.  
  Several studies have addressed the question of whether the interfacial region excerts an own resistance to mass transfer \citep{stephan2019interfacial,enders2008interfacial,nagl2020interfacial,schaefer2023mass}. Some studies report a relationship between interfacial enrichment \citep{li2019molecular,fouad2017phase,llovell2012application,falls1983adsorption,mique2011simultaneous,stephan2019interfacial,klink2014density,klink2015density,garrido2016understanding,karlsson2017dilution,stephan2020enrichment} and the time required for a liquid-liquid system exposed to a chemical potential gradient to equilibrate \citep{deuerling2024investigation,nagl2022interfacial}.
  Similarly, experimental and theoretical studies of evaporation phenomena have consistently reported a temperature jump across the interface \citep{fang1999temperature,ward2001interfacial,badam2007experimental,stierle2020selection}, suggesting that the interface itself has a resistance to heat transfer.
  
  Because directly accessing the structure of interfaces on the molecular scale through experimental methods is difficult, most studies on interfacial properties use theoretical approaches. Molecular simulations, particularly molecular dynamics (MD) simulations, provide detailed predictions of interfacial properties.
  Many studies use MD to investigate  heat transfer in interfacial systems, which is often accompanied by mass transfer, e.g.\ in the case of evaporation \citep{heinen2016communication,homes2021evaporation,lofti2014evaporation,heinen2019nonequilibrium,tsuruta1999condensation,cheng2011evaporation,simon2006interface,ge2007transfer,matsumoto1996molecular,kryukov2011about,kon2017kinetic,kjelstrup1996nonequilibrium,rosjorde2000nonequilibrium,rosjorde2001nonequilibrium,hafskjold1996non,ge2007integral,bird2019transport,inzoli2011transfer,vanderHam2010modelling,chatwell2019diffusion}. 
  Further studies employ theoretical approaches, such as classical density functional theory (DFT) and integral theories, to determine heat transfer resistances within interfaces \citep{johanessen2008nonequilibrium,johanessen2006integral,glavatskiy2010transport,klink2015analysis}. 
  MD simulations have also been applied to study stationary mass transfer phenomena, e.g.\ in pores, membrane, and crystals \citep{lisal2004dual,arya2001critical,hato2012simulation,martin2001effect,pohl1996molecular,pohl1999massively,macelroy1994nonequilibrium} and in homogeneous fluid phases \citep{thompson1998direct,thompson1999direct,sun2007temperature}.
  These studies utilize variants of the dual control volume approach \citep{heffelfinger1994diffusion,heffelfinger1998massively,ford1998massively}, whereby a stationary molecular flux is induced by assigning different chemical potentials to two control volumes. 
  Similar approaches have been used to study the stationary mass transfer across vapor-liquid interfaces \citep{stephan2021mass,rauscher2022nonequilibrium}. 
  
  Various MD simulation approaches have been used to study non-stationary mass transfer at interfaces, including abruptly creating non-equilibrium conditions by deleting particles at the interface \citep{baidakov2019molecular,baidakov2019relaxation} or volume expansion \citep{bucior2009molecular}, as well as tracking individual molecules crossing vapor-liquid interfaces \citep{braga2016nonequilibrium,braga2014nonequilibrium,garrett2006molecular}.
  Recently, \citet{schaefer2023mass} and \citet{braten2023molecular} studied non-stationary mass transfer across vapor-liquid interfaces through the localized insertion of a second component into the vapor phase of an initially equilibrated vapor-liquid system.  
  
  While these studies provide detailed insights into mass transfer, MD simulations require long simulation times or multiple system replicas \citep{casalino2020beyond,zhang2015reliable} to achieve a sufficient signal-to-noise ratio. 
  More importantly, coupling MD calculations of discrete particles or molecules to continuum mechanics models requires averaging noisy signals, whereas DFT natively provides noise-free, averaged quantities. 
  Therefore, alternative approaches based on average quantities rather than atomic/molecular coordinates were applied to investigate mass transfer at interfaces, as these approaches offer a lower computational cost.
  One such approach is dynamic concentration gradient theory \citep{kruber2017interfacial,chicaroux2019theoretical}, which is an incompressible version of density gradient theory extended to the dynamic case. Density gradient theory is based on the Cahn-Hilliard equation \citep{cahn1958free} and uses a functional for the Helmholtz energy of an inhomogeneous system that depends on the local density and its gradients. Coupled with an equation of state, density gradient theory allows for the calculation of equilibrium density profiles and surface tensions of fluid-fluid systems  \citep{breure2012modeling,oliveira2008surface,liang2016density}. Dynamic concentration gradient theory provides differential equations for the time evolution of the concentration of each component in a multicomponent mixture \citep{kruber2017interfacial,chicaroux2019theoretical}. This approach has been applied to mass transfer across liquid-liquid interfaces in systems with up to four components  \citep{deuerling2024investigation,nagl2022study,nagl2022interfacial,nagl2020interfacial,chicaroux2019theoretical,kruber2017interfacial}. 
  
  However, dynamic concentration gradient theory requires experimental data on surface tension and mass transfer in order to adjust influence and mobility parameters, respectively. Additionally, this approach lacks a momentum balance; therefore, it only captures diffusive effects and is limited to incompressible systems such as liquid-liquid interfaces.  
  Another approach based on the Cahn-Hilliard equations is the Cahn-Hilliard/Navier-Stokes model. This model includes a momentum balance and has been applied to study droplet coalescence \citep{zimmermann2020calculation}. 
  In compressible systems, the Navier-Stokes-Korteweg equations incorporate the Korteweg tensor, a constitutive relation involving density gradients, into the momentum balance to capture the effects of vapor-liquid interfacial tension \citep{korteweg1901sur,dunn1985thermomechanics,heida2010compressible,diehl2016numerical,rohde2005local}.
  A similar model has been applied to derive  a description of  heat and mass transfer at vapor-liquid interfaces that does not require the resolution of density and temperature profiles across the interface  \citep{glavatskiy2013nonequilibrium}.
  Helmholtz energy functionals based on the Cahn-Hilliard equation depend on local density values and their gradients. While this approach is sufficient for fluid-fluid interfaces, it cannot model molecular layering, which typically appears at solid-fluid interfaces.
  
  An alternative approach is dynamic density functional theory (DDFT), first derived by \citet{marconi1999dynamic,marconi2000dynamic}. It provides time evolution equations for the density and, in some models, the momentum density, thereby extending equilibrium DFT to non-equilibrium systems \citep{teVrugt2020classical}. 
  Unlike DFT, DDFT is not an exact theory because it employs the \emph{adiabatic approximation}. This approximation assumes that the two-body spatial correlation function in a non-equilibrium system is equal to that in an equilibrium system with the same density profile \citep{marconi1999dynamic,marconi2000dynamic,archer2009dynamical,teVrugt2020classical}. Due to this assumption, Helmholtz energy functionals (equilibrium properties) can be used in DDFT \citep{stierle2021hydrodynamic}. Consequently, the functionals are independent of the system's history and memory effects cannot be described. This limitation can be overcome by using the exact power functional theory \citep{schmidt2013power,brader2015power,deLasHeras2018velocity,schmidt2018power}. 
  
  While the initial versions of DDFT were formulated for colloidal systems \citep{marconi1999dynamic,marconi2000dynamic}, \citet{archer2009dynamical} developed a DDFT for molecular fluids, including a momentum balance.
  The present work examines the extension of molecular DDFT to mixtures, which was proposed by \citet{stierle2021hydrodynamic} as hydrodynamic DFT and applied in a similar form by \citet{nold2024hydrodynamic}. It consists of equations describing the time evolution of mass, partial density and momentum, and includes inertia as well as viscous dissipation. 
  This approach has been combined with Maxwell-Stefan diffusion \citep{curtiss1999multicomponent,curtiss2001multicomponent,bird2013multicomponent,stierle2021hydrodynamic} to provide a qualitative description of the coalescence of droplets and bubbles. 
  Hydrodynamic DFT can be combined with non-local Helmholtz energy functionals. Functionals based on the PC-SAFT model \citep{gross2001perturbed,gross2002modeling,gross2003modeling,sauer2017classical,stierle2020guide}, in particular, show significant predictive capabilities for equilibrium properties when compared to experimental data \citep{gross2009density,sauer2017classical,sauer2018prediction,sauer2019prediction,rehner2018surface,rehner2021surfactant,nitzke2023,bursik2024predicting,stierle2024classical,bursik2025staticcontactanglesmixtures}.
  Hydrodynamic DFT with a PC-SAFT-based functional  was applied to study wetting on the microscopic scale, achieving good agreement with non-equilibrium MD (NEMD) simulations \citep{bursik2025modelling}. 
  
  In this work, we assess the predictive capabilities of hydrodynamic DFT for mass transfer across a vapor-liquid interface. Specifically, we apply it to the scenario proposed by \citet{schaefer2023mass}, in which molecules of a second component are inserted into the vapor phase of an initially pure, equilibrated vapor-liquid system. 
  We observe the system's response in certain measurement volumes (MV) close to the interface, as well as by providing density and flux profiles across the interface.
  We combine hydrodynamic DFT for mixtures with Helmholtz energy functionals based on the PeTS model \citep{heier2018equation}, which accurately describes the Lennard--Jones truncated and shifted (LJTS) fluid studied here. We model the viscous shear stress assuming a Newtonian pressure tensor, and employ the Maxwell-Stefan equations for the diffusive molecular fluxes. We determine transport properties using a generalized version of the entropy scaling approach\citep{bursik2024viscosities,stierle2021hydrodynamic}. 
  All results are validated against NEMD simulations for the LJTS fluid. 
  
  This manuscript is structured as follows: In \cref{sec:methods}, we introduce the setup of the vapor-liquid system and classical as well as hydrodynamic DFT. We detail generalized entropy scaling for transport properties, the implementation of  hydrodynamic DFT, and NEMD simulations. 
  Finally, in \cref{sec:results} we present results for the equilibrated vapor-liquid interface, comparing them with MD simulations, as well as with density and molecular fluxes close to and across the interface after the thermodynamic equilibrium is perturbed by the insertion of the second component.

  \section{Methods } \label{sec:methods}
  
  \subsection{Setup of the Vapor-Liquid Test System} \label{sec:methods_setup}
  
  \begin{figure}[ht]
    \centering
    \includegraphics[width=0.7\columnwidth]{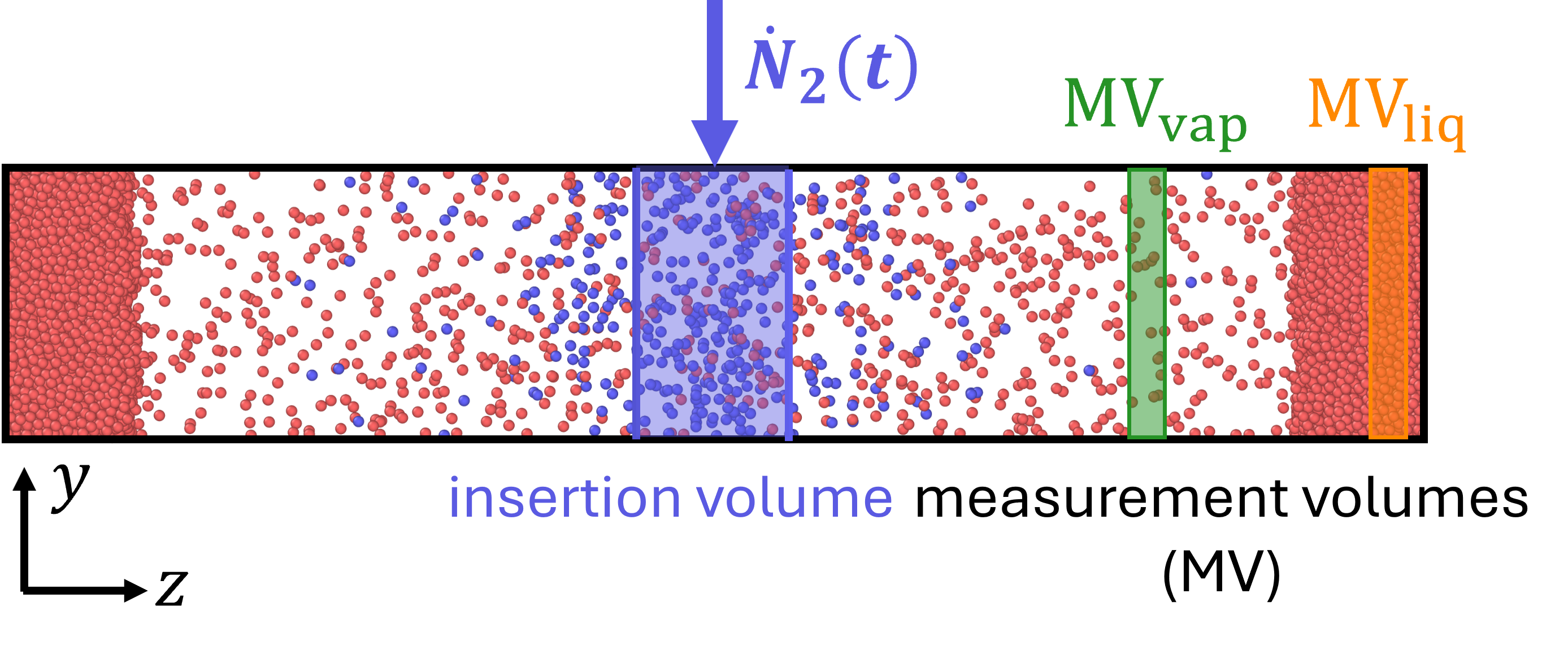}
    \caption{Setup of the system showing the insertion of molecules of component 2 and the measurement volumes in the vapor (MV$_\mathrm{vap}$) and liquid (MV$_\mathrm{liq}$) phase. }
    \label{fig:system}
  \end{figure}
  The mass transfer across the vapor-liquid interface is studied in the system shown in \cref{fig:system}, which is similar to the one used by \citet{schaefer2023mass} and contains an LJTS fluid. Initially, the system contains only component 1 and is at equilibrium. Then, component 2 is inserted at a constant rate, $\dot{N}_2$, within an insertion period, $t_0\leq t \leq t_\mathrm{end}$, resulting in a flux of component 2 in both directions of the $z$-coordinate and across the vapor-liquid interface. 
  To observe the dynamics o the system, two measurement volumes (MV) are defined: one in the vapor phase (MV$_\mathrm{vap}$) and one in the liquid phase (MV$_\mathrm{liq}$). The system's response is observed in these regions. The measurement volumes are placed near the interface because this study focuses on interfacial mass transfer, making these regions particularly relevant.
  
  \begin{figure}[ht]
    \centering
    \includegraphics[width=0.6\columnwidth]{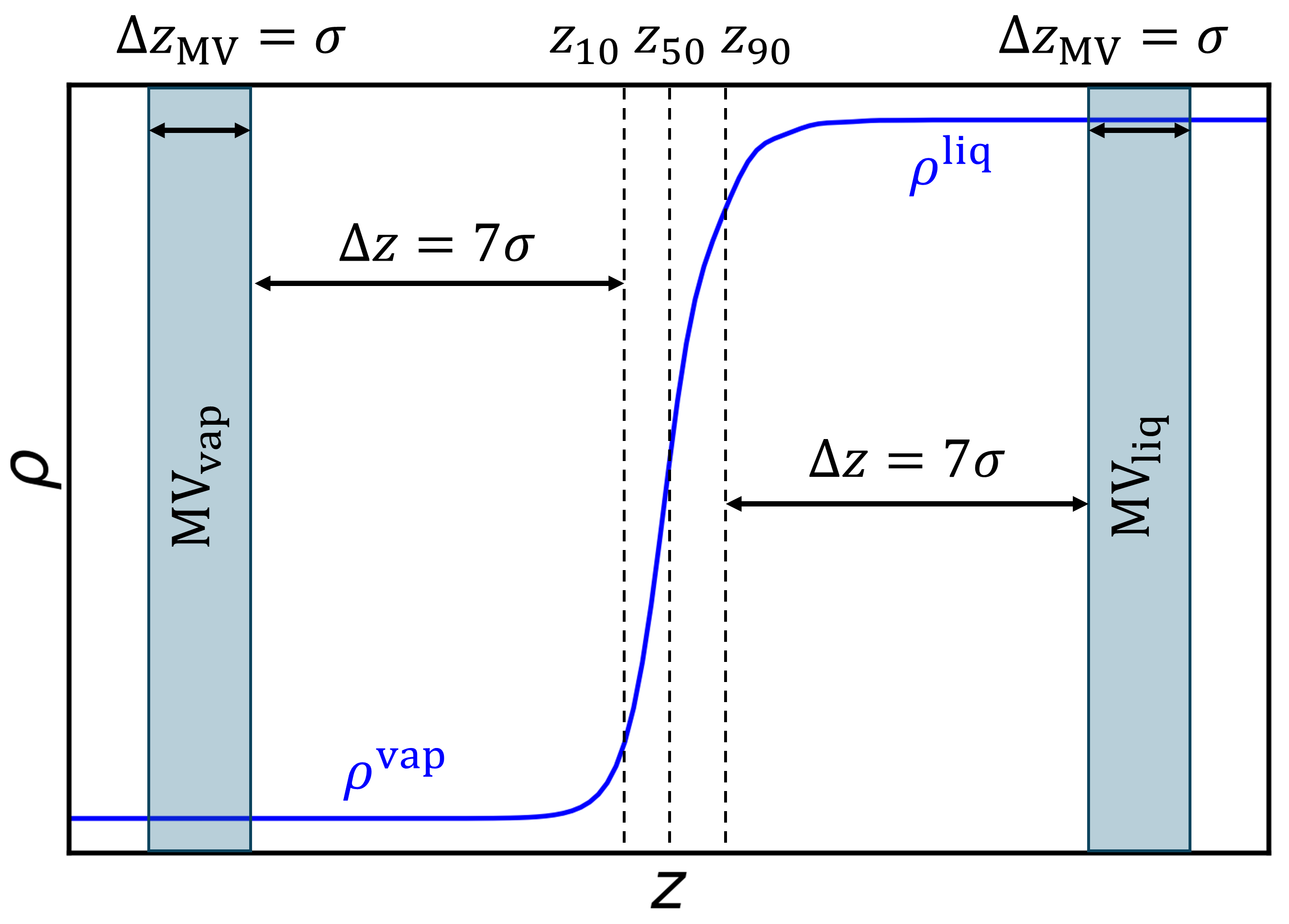}
    \caption{Qualitative density profile of the interfacial region visualizing the $z_{10}$,$z_{50}$ and $z_{90}$ positions close to the interface, as well as the location of the measurement volumes.}
    \label{fig:interface}
  \end{figure}
  \Cref{fig:interface} provides a closer look at the interfacial region.  Three positions at the interface are defined by the $z$-coordinates: $z_{10}=z(\rho_{10})$, $z_{50}=z(\rho_{50})$, and $z_{90}=z(\rho_{90})$ are defined as the $z$ coordinates, where 10\%, 50\%, or 90\% of the difference between the vapor density, $\rho^\mathrm{vap}$, and the liquid density, $\rho^\mathrm{liq}$, is reached, according to 
  \begin{subequations}
    \begin{align}
      \rho_{10} =&\rho^\mathrm{vap} + 0.1 (\rho^\mathrm{liq}-\rho^\mathrm{vap})\\
      \rho_{50} =&\rho^\mathrm{vap} + 0.5 (\rho^\mathrm{liq}-\rho^\mathrm{vap})\\
      \rho_{90} =&\rho^\mathrm{vap} + 0.9 (\rho^\mathrm{liq}-\rho^\mathrm{vap})
    \end{align} 
  \end{subequations}
  The vapor and liquid measurement volumes MV$_\mathrm{vap}$ and MV$_\mathrm{liq}$ are located at $z_{10} - \Delta z$ and $z_{90} + \Delta z$, respectively, where $\Delta z=7\sigma$ and $\Delta z_\mathrm{MV}=\sigma$. All positions are recalculated for each time step because the density at the interface fluctuates statistically (only for NEMD) and due to the insertion of component 2 (for both hydrodynamic DFT and NEMD).
  
  \subsection{Classical Density Functional Theory} 
  
  This section introduces the DFT approach employed in this work to describe inhomogeneous (vapor-liquid) equilibrium systems, building upon previous work \citep{gross2009density,klink2014density,sauer2017classical,sauer2018prediction,sauer2019prediction}. DFT is an exact equilibrium theory formulated as a variational principle \citep{evans1979nature,evans2016new} based on the grand potential functional $\Omega$ providing equilibrium density profiles $\rho_i(\rb)$ of each component $i$ in a mixture of $N$ components. Given the chemical potentials $\mu_i$, the volume $V$, and the temperature $T$, the grand potential functional is defined as 
  \begin{equation} \label{eq:Omega}
    \Omega\left[\{\rho_i(\rb)\}\right] =F\left[\{\rho_i(\rb)\}\right]  -\sum_{i=1}^{N} \int \rho_i(\rb) \left(\mu_i-V_i^\ext(\rb)\right)\dr
  \end{equation}
  where $F$ is the Helmholtz energy functional that models fluid-fluid interactions in the system. Square brackets indicate a functional dependence, and curly brackets denote dependence on the set of partial densities, $\rho_i$, of all components. The external potential, $V_i^\ext$, captures solid-fluid interactions, which are not required for the present work, so that $V_i^\ext=0$.
  
  In equilibrium, the grand potential functional attains a minimum with respect to the set of density profiles, $\{\rho_i(\rb)\}$, and its functional derivatives vanish, according to 
  \begin{equation}
    \left. \frac{\delta\Omega}{\delta\rho_i} \right|_{\{\rho_i^\mathrm{eq}(\rb)\}} =0
  \end{equation}
  This leads to the Euler-Lagrange equation 
  \begin{equation} \label{eq:ELE}
    \frac{\delta F}{\delta \rho_i(\rb)}-\mu_i+V_i^\ext(\rb)=0 \qquad\qquad \forall i
  \end{equation}
  from which the equilibrium density profiles, $\rho_i^\mathrm{eq}(\rb)$, can be obtained. For the remainder of this work, we will drop the superscript `eq' and differentiate between dynamic and equilibrium densities by other means. 
  Up to this point, DFT is an exact equilibrium theory that does not rely on approximations \citep{evans1979nature,evans2016new}. However, the Helmholtz energy functionals, $F$, are not exactly known for most fluids and must be modeled.  
  
  In this work, we use the Helmholtz energy functional based on the PeTS equation of state \citep{heier2018equation} which is a reparametrized PC-SAFT \citep{gross2001perturbed,gross2002modeling,gross2003modeling} model for the LJTS fluid. It consists of additive contributions according to 
  \begin{equation}
    F  =F^\mathrm{ig}+F^\mathrm{hs} +F^\mathrm{disp}
  \end{equation}
  where $F^\mathrm{ig}$, $F^\mathrm{hs}$, and $F^\mathrm{disp}$ are the functional contributions for the ideal gas, hard-sphere fluid, and dispersive interactions, respectively. 
  The expression for $F^\mathrm{ig}$ is known exactly from statistical mechanics. 
  For the hard-sphere contribution, the White Bear version of fundamental measure theory (FMT) \citep{yu2002structures, roth2002fundamental} is employed. This formulation is consistent with the Boublík-Mansoori-Carnahan-Starling-Leland (BMCSL) equation of state \citep{mansoori1971equilibrium, boublik1970hard}.
  The functional for dispersive interactions is modeled using a weighted density approximation (WDA), as described in \citet{sauer2017classical} and  adopted by \citet{heier2018equation} 
  \begin{equation}
    \beta F^{\mathrm{disp}}\left[ \{ \rho_i(\rb) \} \right] = \int \bar{\rho}^{\mathrm{disp}}(\rb) \tilde{a}^{\mathrm{disp}}(\{\bar{\rho}_i^{\mathrm{disp}}(\rb)\}) \dr
  \label{eq:DispersionFunctional}
  \end{equation}
  where $\tilde{a}^{\mathrm{disp}}$ is the dispersive contribution to the  Helmholtz energy (per molecule) for the PeTS equation of state. The weighted density for the dispersive contribution is given by
  \begin{equation}
      \bar{\rho}_i^{\mathrm{disp}}(\rb) = \frac{3}{4 \pi \left(\psi^\mathrm{disp} d_i\right)^3} \int \rho_i(\rb') \Theta\left( \psi^\mathrm{disp} d_i - |\rb - \rb'| \right) \dr' 
      \label{eq:WeightedDensityDISP}
    \end{equation}
  with the Heaviside step function $\Theta$ and the universal parameter $\psi^\mathrm{disp} = 1.21$, which affects the range of averaging by the integral. 
  This model parameter was fitted to molecular simulation data for surface tensions of the LJTS fluid \citep{heier2018equation}. 
  $d_i(T) $ is the temperature-dependent, effective hard-sphere diameter \citep{gross2001perturbed,heier2018equation} according to 
  \begin{equation}
    d_i(T)= \sigma_\mathrm{ii} \left(1-c_1\exp\left(-c_2\frac{\varepsilon_\mathrm{ii}}{k_\mathrm{B}T}\right) \right)
  \end{equation}
  with the parameters $c_1=0.127112544$ and $c_2 = 3.052785558$ according to \citet{heier2018equation}.
  This Helmholtz energy functional has two model parameters defining the fluid: the energy interaction parameter, $\varepsilon_{ii}$, and the diameter parameter, $\sigma_{ii}$, for each component $i$. Interactions between unlike molecules are modeled by modified Lorentz-Berthelot combining rules, according to 
  \begin{equation} \label{eq:combining}
    \begin{aligned}
      \sigma_{ ij}      & = \frac{\left(\sigma_{ii} + \sigma_{jj}\right)}{2} \\
      \varepsilon_{ ij} & = \sqrt{\varepsilon_{ii} \varepsilon_{jj}}(1-k_{ij})
    \end{aligned}
    \end{equation}
  where a binary parameter, $k_{ij} \neq 0$, can be used to define the cross energy parameter in the mixture. The same combining rules are employed in the MD simulations.

  \subsection{Hydrodynamic Density Functional Theory}
  
  The hydrodynamic DFT model for mixtures, originally developed for pure fluids\citep{archer2009dynamical}, is taken from \citet{stierle2021hydrodynamic}. The model consists of mass, component, and momentum balances, according to
  \begin{subequations}
    \begin{align}
      \frac{\partial \left(\mw\rho\right)}{\partial t} + \divd (\mw\rho \vb) &=                                 0  \label{eq:ContinuityDDFT}                                                                                                                                            \\
      \frac{\partial \rho_i}{\partial t} + \divd (\rho_i \vb + \jdiff) &= 0   \label{eq:ComponentDDFT}                                                                                                                                            \\
      \frac{\partial (\mw\rho \vb)}{\partial t} + \divd \left(\mw\rho \vb \vb^\intercal + \boldsymbol{\tau}\right) &=  - \sum_i^N \rho_i \nabla  \left( \frac{\delta  F}{\delta \rho_i} +  V^\ext_i \right)   \label{eq:MomentumDDFT}
    \end{align}
    \label{eq:DDFTModel}
  \end{subequations}
  where $\jdiff \equiv \rho_i(\vb_i - \vb)$  is the molecular flux relative to  the barycentric (mass-averaged) velocity $\vb$. Furthermore, $\mw$ is the average molecular mass in the mixture, and $\boldsymbol{\tau}$ is the shear pressure tensor. The component balances are linearly dependent because the overall mass balance is obtained by summing all the component balances. Note that the sum of the molecular fluxes vanishes, or equivalently, 
  \begin{equation} \label{eq:sum_jdiff}
    \mw_N \j_N^\diff = -\sum_i^{N-1} \mw_i\jdiff
  \end{equation}
  
  Hydrodynamic DFT has been shown to accurately capture the effects of interfaces and external fields (which are not relevant here; $V^\ext_i=0$) on the dynamics of the system 
  \citep{bursik2025modelling}.
  This is achieved by incorporating the DFT term, $- \sum_i^N \rho_i \nabla \frac{\delta  F}{\delta \rho_i}$, into the momentum balance, which predicts the influence of fluid-fluid interfaces without requiring information about interfacial tensions. 
  
  The shear pressure tensor is modeled under the assumption of a Newtonian fluid (with negligible bulk viscosity) as
  \begin{equation*}
    \boldsymbol{\tau} = - \eta \left( \nabla\vb + (\nabla\vb)^\intercal - \frac{2}{3} \left( \divd \vb \right) \mathbb{I} \right)  \label{eq:tau}
  \end{equation*}
  with the shear viscosity, $\eta$, of the mixture. Although the shear viscosity of inhomogeneous systems generally has tensorial rank, we assume it is a scalar value. This assumption has been shown to be reasonable for interfacial systems \citep{bursik2025modelling}. However, this scalar value is considered to depend on position within the system; that is, local values of shear viscosity are used.
  
  The molecular fluxes are described using generalized Maxwell-Stefan diffusion. This approach was used in a study of droplet and bubble coalescence \citep{stierle2021hydrodynamic}. It can be written as follows  
    \begin{equation}
      - \frac{\rho_i \nabla \mu_i^\diff}{\kb T} = \sum_{j\neq i}\frac{1}{D_{ij}}\left(x_j  \mathbf{j}_i^\diff - x_i  \mathbf{j}_j^\diff \right) \label{eq:MS}
      \end{equation}
  where $D_{ij}$ is the Maxwell-Stefan diffusion coefficient and $x_i$ is the mole fraction of component $i$. Similar to viscosity, the Maxwell-Stefan diffusion coefficient is assumed to depend on position within the inhomogeneous system, requiring local values. 
  The generalized driving force for Maxwell-Stefan diffusion $\nabla \mu_i^\diff$ \citep{curtiss1999multicomponent,curtiss2001multicomponent,bird2013multicomponent} was extended to inhomogeneous systems by \citet{stierle2021hydrodynamic} and is defined as
      \begin{equation}
        \rho_i \nabla \mu_i^\diff = \rho_i\nabla \left(\frac{\delta F}{\delta\rho_i} + V_i^\ext\right)   -w_i\sum_j \rho_j \nabla\left(\frac{\delta F}{\delta\rho_j} + V_j^\ext\right) \label{eq:MS}
      \end{equation}
  with the mass fraction $w_i$ of component $i$. 
  The first term in this definition of the generalized driving force for diffusion represents the gradient of the total chemical potential (according to \cref{eq:ELE}), which accounts for the presence of interfaces and external fields. The second term ensures that the sum of all driving forces vanishes, in agreement with \cref{eq:sum_jdiff}  \citep{curtiss1999multicomponent,curtiss2001multicomponent,bird2013multicomponent,stierle2021hydrodynamic}.
  
  \subsection{Transport Properties from Generalized Entropy Scaling }
  
  In the hydrodynamic DFT model, the shear viscosity, $\eta$, and the Maxwell-Stefan diffusion coefficient, $D_{ij}$, depend on position within the system, especially at interfaces. Local values are determined using generalized entropy scaling \citep{bursik2024viscosities}, which extends the entropy scaling approach for viscosity \citep{rosenfeld1977relation, rosenfeld1999quasi} to inhomogeneous systems. In this work, a similar approach is used for the self-diffusion coefficient, from which the Maxwell-Stefan diffusion coefficient can be estimated. 
  The entropy scaling approach utilizes the univariate relationship between transport properties and residual molar entropy \citep{loetgeringlin2015group,loetgeringlin2018pure}. 
  
  In the inhomogeneous case, the local values of the residual entropy density (i.e.\ entropy per unit volume) are determined from
  \begin{equation} \label{eq:sres_density}
    \tilde{s}_\mathrm{res}(\rb) = -\left(\frac{\partial f_\mathrm{res}(\rb)}{\partial T}\right)_{\rho,V}
  \end{equation}
  where $f_\res$ is the residual Helmholtz energy density. Note that $f_\res$ is also used in hydrodynamic DFT to calculate the DFT term $-\sum_i^N \rho_i \nabla \frac{\delta  F}{\delta \rho_i}$ with $F^\mathrm{res} = \int f^\mathrm{res}\d \rb$, rendering the generalized entropy scaling approach consistent with hydrodynamic DFT regarding the molecular model. 
  The dimensionless residual entropy profile (per molecule) is defined as \citep{bursik2024viscosities}
  \begin{equation} \label{eq:sres}
    s_\mathrm{res}^{\#}(\rb) = \frac{\tilde{s}_\mathrm{res}(\rb)}{\bar{\rho}^\mathrm{ES}(\rb)k_\mathrm{B}}
  \end{equation}
  with the weighted density, $\bar{\rho}^\mathrm{ES}(\rb)$, which considers non-local effects, i.e.\ the influence of interfaces. For a liquid-vapor system, it is defined as
  \begin{equation} \label{eq:rho_bar_ES_tildepsi}
    \bar{\rho}^\mathrm{ES}(\rb) = \sum_i \frac{3}{4\pi\left(\psi^\mathrm{ES} d_i\right)^3} \int \rho_i(\rb) \Theta(\psi^\mathrm{ES}  d_i - |\rb-\rb'|) \d \rb
  \end{equation}
  An adjustable parameter, $\psi^\mathrm{ES}$, can capture the effect of interfaces on transport properties \citep{bursik2024viscosities,bursik2025modelling}; it is set to unity, $\psi^\mathrm{ES}=1$ in this work.
  Viscosity and self-diffusion coefficient profiles can be determined from the dimensionless residual entropy profile (\cref{eq:sres}) by evaluating the ansatz functions of the homogeneous systems at each position. The ansatz function and the procedure for estimating Maxwell-Stefan diffusion coefficients from self-diffusion coefficients are provided in \cref{sec:appendix_entropyScaling}.

  \subsection{Implementation of Hydrodynamic DFT}
  
  Spatial discretization is performed using a first-order, well-balanced, finite-volume scheme, as described by \citet{stierle2021hydrodynamic} and developed by \citet{carrillo2021high}. 
  This approach preserves the stationary states of the original system and ensures that the sum of kinetic, Helmholtz, and potential energy decreases monotonically over time.
  
  \citet{stierle2024classical}  implemented Helmholtz energy functionals and equilibrium DFT using JAX \citep{jax2018github}, which facilitates the efficient application of automatic differentiation and GPU parallelization by leveraging algorithmic advances in machine learning.
  Building on their work, we extended the software to include hydrodynamic DFT equations for multicomponent systems.
  For time integration, we use the Diffrax library \citep{kidger2022neuraldifferentialequations}, along with a custom implementation of the Runge-Kutta method SSPERK(6,4) \citep{fekete2022}, due to its large stability region. This is beneficial for stiff systems, such as the hydrodynamic DFT equations. 
  The embedded Runge-Kutta pair enables adaptive step size control based on error estimation.
  This approach uses end-to-end just-in-time compilation to efficiently integrate the hydrodynamic DFT equations on GPUs in single- or multidimensional domains with low overhead. This leads to substantial speedups compared to previous implementations.
      
  The hydrodynamic DFT simulations in this work are performed in a one-dimensional domain with periodic boundary conditions. The domain has a length of $L_z = 150\sigma$ and contains 2048 grid cells. A source term is introduced into the component balance (\cref{eq:ComponentDDFT}) to add  molecules of component 2 during the period of molecule insertion. For the DFT calculations, an initial profile is chosen such that the desired number of molecules of each component is present in the system (consistent with the MD simulations). Then, a mathematical reformulation is employed to solve the DFT equations while keeping the number of molecules of each component constant \citep{rehner2018surface,bursik2025modelling}. 
  Because the system is sufficiently diffusive, numerical diffusion in the Lax-Friedrichs scheme for convective fluxes could be omitted in favor of a more accurate solution. 
  
  \subsection{Non-Equilibrium Molecular Dynamics Simulations}

  The results from hydrodynamic DFT are compared to NEMD simulations. The simulations were performed using the Large-scale Atomic/Molecular Massively Parallel Simulator (LAMMPS, stable release 2 Aug 2023) \citep{thompson2022lammps}, where minor adjustments to the source code were necessary to insert particles with a velocity distribution consistent with the desired temperature. 
  The general setup of the system was described in \cref{sec:methods_setup} 
   above, and it is defined similar to the work of \citet{schaefer2023mass}. 
  We conducted our own MD simulations after observing a slight initial broadening of the interface in the previously reported data, a behavior we did not anticipate and were unable to reproduce in our MD simulations. Although the effect is minor, it motivated us to independently verify the results.
  The simulations consist of three distinct steps: (i) Equilibrium MD simulations are performed for a system containing only component 1 molecules. (ii) During the insertion phase, molecules of component 2 are inserted. (iii) The response of the system is simulated during and after the insertion phase. Steps two and three are performed using NEMD simulations. Step one (equilibration) and three (simulation after the insertion period) are carried out in the canonical ensemble, i.e., at constant $N,V$, and $T$. Although the insertion rate is prescribed directly in LAMMPS, this procedure is equivalent to setting a constant chemical potential of component~2, $\mu_2$, in the insertion volume. 
  
  The molecular interactions in the simulations are modeled using the LJTS molecular model, in which molecules interact according to \citep{allen1989computer}
  \begin{subequations}
    \begin{align}
      &u_{ij}^{\mathrm{LJ}}(r)=4 \varepsilon_{ij}\left[\left(\frac{\sigma_{ij}}{r}\right)^{12}-\left(\frac{\sigma_{ij}}{r}\right)^6\right] \\
      &u_{ij}^{\mathrm{LJTS}}(r)= \begin{cases}u_{ij}^{\mathrm{LJ}}(r)-u_{ij}^{\mathrm{LJ}}\left(r_{\mathrm{c}}\right) & r \leq 2.5 \sigma_{ij} \\ 0 & r>2.5 \sigma_{ij}\end{cases}
      \end{align}
  \end{subequations}
  with the Lennard--Jones potential, $u_{ij}^{\mathrm{LJ}}$, depending on the distance, $r$, between two interacting particles, as well as the Lennard--Jones energy parameter, $\varepsilon_{ii}$, diameter, $\sigma_{ii}$, and particle mass, $m_{i}$. We choose a cutoff radius of $r^*_{\mathrm{c}}=2.5$, at which the potential of the LJTS fluid $u_{ij}^{\mathrm{LJTS}}$ is truncated and shifted to zero. 
  The interaction of unlike molecules is determined from the Lorentz-Berthelot combining rules, analogously to the hydrodynamic DFT approach (cf.\ \cref{eq:combining}).

  The equations of motion are integrated using the velocity-Verlet method throughout the three steps of the simulation. The time step for the initial equilibration is set to $\Delta t^*=0.005t^*$, where $t^*$ is the dimensionless time.  The asterisk~$*$ indicates dimensionless variables, which are used for all quantities in this work.
  During and after insertion, the time step is reduced to $\Delta t^*=0.001t^*$. The Lennard--Jones parameters ($\varepsilon_{11}$, $\sigma_{11}$, and $m_1$) are used to calculate the dimensionless variables provided in \cref{table:dimensionless}.

  \begin{table}[h!]
    \centering
    \caption{Dimensionless quantities, as denoted by an asterisk~$*$, are reduced using the Lennard--Jones parameters $\sigma_{11}$, $\varepsilon_{11}$ and $m_{1}$.}
    \label{table:dimensionless}
    \renewcommand{\arraystretch}{1.2}
    \begin{tabular}{r c  l@{\hskip 0.5cm} l@{\hskip 0.0cm}  }
        \hline
        $z^*$                       & $=$ & $ z /\sigma_{11}$                                                              & $z$-position                      \\
        $\rho^*$                    & $=$ & $ \rho \sigma_{11}^3$                                                          & density                         \\
        $v^* $                      & $=$ & $  v/\sqrt{\varepsilon_{11} / m_{1}}$                           & velocity                        \\
        $T^*$                       & $=$ & $ \frac{T}{\varepsilon_{11} / k_{\mathrm{B}}}$                                 & temperature                     \\
        $\eta^*$                    & $=$ & $ \eta\sigma_{11}^2/\sqrt{\varepsilon_{11} m_{1}}$                & shear viscosity                 \\
        $\D^*$                    & $=$ & $ \frac{D}{\sigma_{11} \sqrt{\varepsilon_{11} / m_{1}}}$                & diffusion coefficient        \\
        $\j^*$                    & $=$ & $ j  \sigma_{11}^3 \sqrt{m_{1}/\varepsilon_{11}}$                & molecular flux        \\
        $t^*$                       & $=$ & $ \frac{t}{\sigma_{11} \sqrt{m_{1} / \varepsilon_{11}}}$ & time                            \\
        \hline
    \end{tabular}
  \end{table}
  
  The temperature is thermostatted using three Nose-Hoover chains \citep{tuckermann2006liouville} with a damping time constant of $t_\mathrm{D}^*=100\Delta t^*$. It is important to note that the inserted molecules of component 2 must be considered in the temperature calculation and thermostatting during and after the insertion phase.  
  To avoid the accumulation of net momentum due to numerical inaccuracies, the net momentum in all directions was removed every 100 time steps. 
  
  The NEMD simulations are carried out in a three-dimensional Cartesian system (cf.\ \cref{fig:system}). The system is $L_z^* = 150$ long and has a quadratic cross section with sides measuring $L_x^*=L_y^*=27.85$. Periodic boundary conditions are employed in all directions, consistent with the hydrodynamic DFT simulations. 
  Initially, the system contains $N_1=16000$ particles. The initial conditions are chosen so that a liquid slab forms at each end of the box in the $z$-direction during equilibration and a vapor phase exists between these two liquid slabs. The densities of the liquid slab and the vapor phase are estimated using the PeTS equation of state. The width of each region is determined based on the estimated density to ensure that $N_1=16000$ particles are introduced into the system. 
  During the period of particle insertion, $N_2=1200$ particles are added to the system in the insertion volume located within the vapor phase at the center of the box. 
  
  The pure system was equilibrated using a conjugate gradient method and \num{2.5e5} time steps. \num{4.75e6} steps were performed during and after the insertion phase. The positions of the molecules were written out every 1000 time steps. To obtain a high signal-to-noise ratio, symmetry was employed, and the results were averaged in the $x$- and $y$-direction to provide one-dimensional profiles of quantities such as the density $\rho(z,t)$. 
  Furthermore, to achieve meaningful signal-to-noise ratios for time-dependent results, where time averaging, as commonly used for steady-state quantities, is not possible, 200 replicas of each system were generated and averaged, as described in \citet{schaefer2023mass}.
  Statistical uncertainties are obtained from the results of different replicas, and all error bars in this work denote the $95\%$ confidence interval. 
  
  One of the main results of the simulations is the molar flux, $j_2$, across the interface, which must be determined in a post-processing step. Following \citet{schaefer2023mass}, the molar flux $j_2$ is calculated by solving the component balance for component 2 in each bin according to 
  \begin{equation}
    \left(\frac{\partial \rho_2}{\partial t}\right)^k = - \frac{j_2^{k+1}-j_2^k}{\Delta z_\mathrm{bin}} 
  \end{equation}
  where the superscript $k=0,...,N_\mathrm{bin}$ indicates the $k$-th bin and the flux $j_2^k$ is the flux entering bin $k$ at its left boundary.
  The $N_\mathrm{bin}+1$ fluxes are obtained by solving the component balance in each of the $N_\mathrm{bin}$ bins and assuming a vanishing flux (symmetry boundary condition) in the middle of the liquid slab. 
  
  \section{Results and discussion} \label{sec:results}
  In the following we present results for the equilibrium and non-equilibrium densities, as well as the molecular fluxes, with a particular focus on the interfacial regions. We assess predictions from (hydrodynamic) DFT by comparing them with results from (NE)MD. For completeness, we provide the phase behavior of the two investigated mixtures. 
  
  \subsection{Phase Behavior of Mixtures}
  
  \begin{figure}
    \centering
  
    \begin{subfigure}[b]{0.475\textwidth}
      \includegraphics[width=\textwidth]{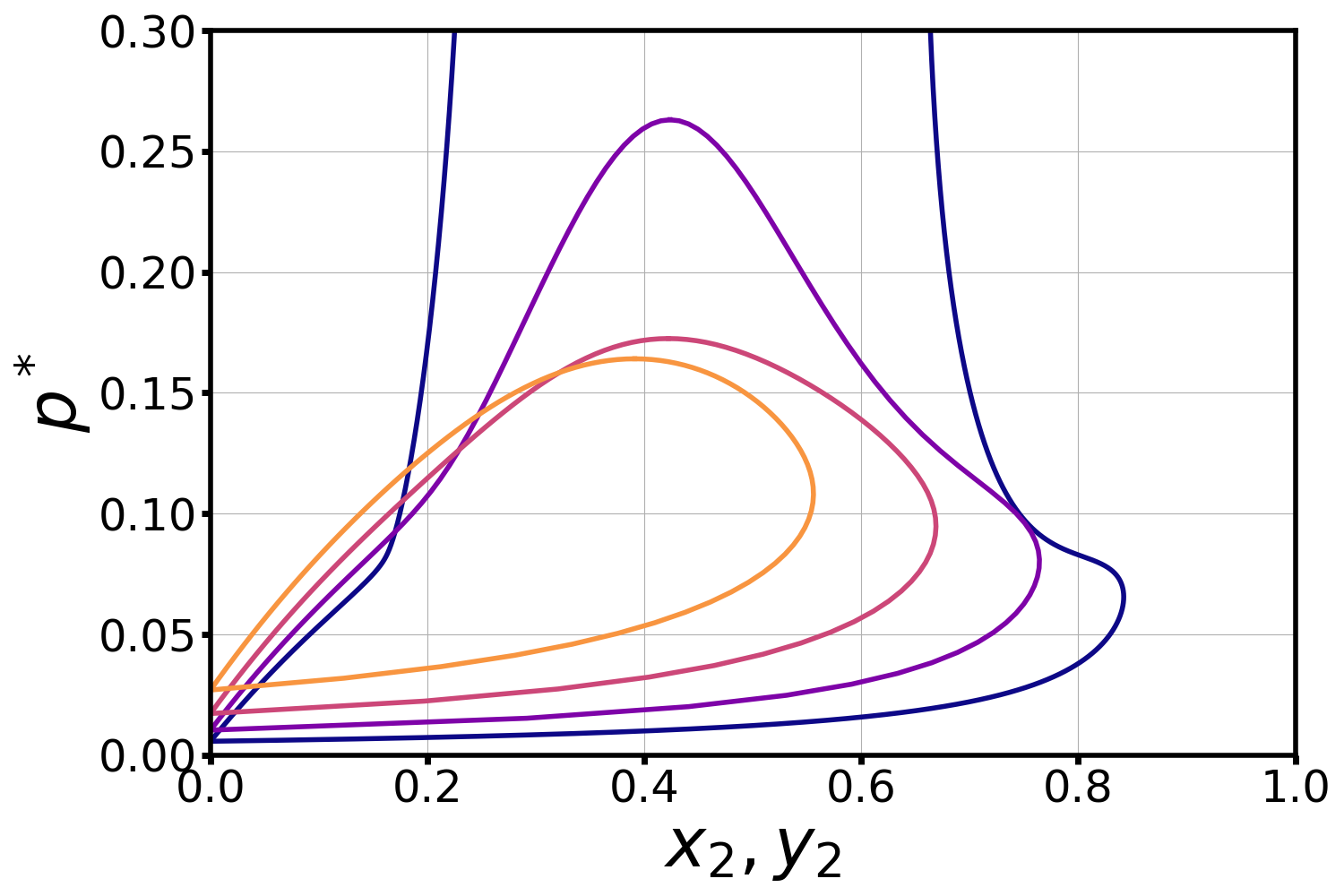}
      \caption{Mixture~A.}
      \label{fig:vle_eos_A}
    \end{subfigure}
    \hfill
    \begin{subfigure}[b]{0.475\textwidth}
      \includegraphics[width=\textwidth]{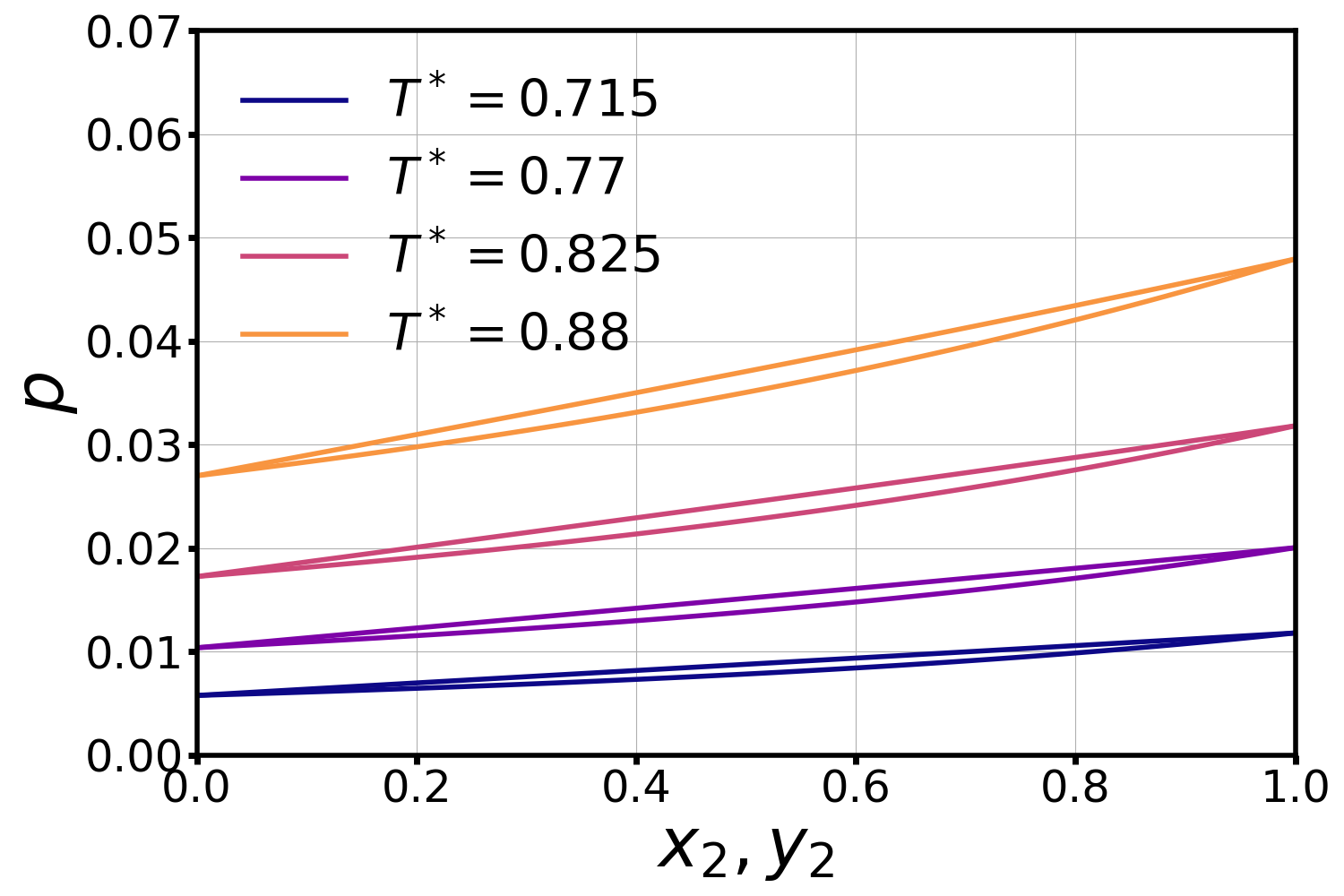}
      \caption{Mixture~B.}
      \label{fig:vle_eos_B}
    \end{subfigure}
  
    \caption{Phase diagrams for the two mixtures for different temperatures $T^*$ determined using the PeTS equation of state.}
    \label{fig:vle_eos}
  \end{figure}
  
  Following \citet{schaefer2023mass}, two significantly different mixtures are chosen in this work, to compare the influence of the type of mixture on mass transfer across the interface. 
  Mixture~A is a highly non-ideal mixture, and mixture~B is close to ideal in the sense of Raoult's law (the bubble point curves depend linearly on composition). This is achieved by altering the energy interaction parameter ratio $\varepsilon_{22}/\varepsilon_{11}$, and by defining a cross-energy parameter $\varepsilon_{ij}$ through the binary interaction parameter, $k_{ij}$, in the combining rules (cf.\  \cref{eq:combining}) as 
  \begin{itemize}
    \item Mixture~A: $\varepsilon_{22}/\varepsilon_{11}=0.6$ and $k_{12}=0.15$
    \item Mixture~B: $\varepsilon_{22}/\varepsilon_{11}=0.9$ and $k_{12}=0.0$
  \end{itemize}
  where the same diameter parameter  $\sigma_{ii}$ is used in both mixtures. 
  According to the literature, the phase behavior of these mixtures is well described by the PeTS equation of state, though there are some deviations in the partial density of the second component \citep{stephan2019interfacial,heier2018equation,schaefer2023mass}.
  
  \Cref{fig:vle_eos} shows the phase equilibria of both mixtures at different temperatures, as determined using the PeTS equation of state.  Mixture~A (cf.\ \cref{fig:vle_eos_A}) exhibits asymmetric, wide-boiling phase behavior.
  This behavior is caused by the strong difference in molecular interactions between the two types of molecules ($\varepsilon_{22}/\varepsilon_{11}=0.6$) and weak interaction between unlike molecules ($k_{12}=0.15$ in \cref{eq:combining}). 
  
  In contrast, for mixture~B (cf.\ \cref{fig:vle_eos_B}), the dew and bubble point lines are located close to each other. The behavior of the mixture is close to that of an ideal mixture, caused by similar molecular interactions between like and unlike molecules. This is consistent with the definition of mixture~B, for which the interaction energy between like and unlike molecules is similar ($\varepsilon_{22}/\varepsilon_{11}=0.9$ and $k_{12}=0$).

  \subsection{Interfaces in Equilibrium}
  
  \begin{figure}[ht]
    \centering
    \includegraphics[width=0.6\columnwidth]{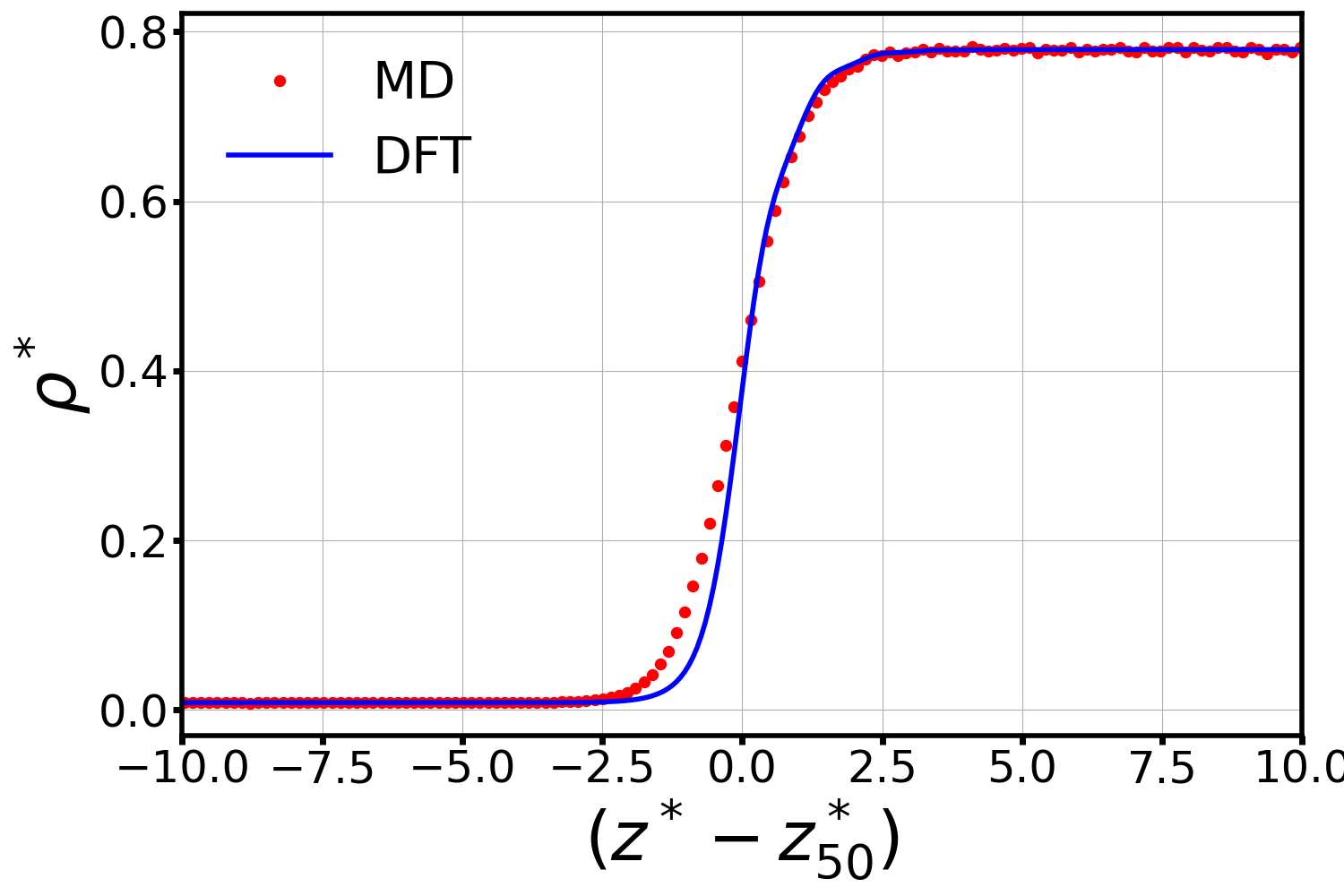}
    \caption{Equilibrium density profile of the pure system containing only component 1 from DFT and MD at $T^*=0.715$. }
    \label{fig:rho_eq_pure}
  \end{figure}
  Before insertion, the system is equilibrated and contains only component 1. Similarly, after the insertion of component 2, the system reaches equilibrium at the end of the simulation. This section discusses density profiles across the interface for both cases. \Cref{fig:rho_eq_pure} shows the equilibrium density profiles for the pure system containing only component 1. The density exhibits a monotonous transition from vapor to liquid density in both DFT and MD. The interface is slightly wider in the MD case, especially in the region where the density first increases from the vapor value. This effect has been observed in previous studies of the LJTS fluid and has been attributed to the movement of the liquid slab and fluctuations in the liquid surface \citep{stephan2018vapor}. Nevertheless, DFT predictions agree well with MD results and can be used as initial profiles for hydrodynamic DFT simulations.
  
  \begin{figure}
    \centering
  
    \begin{subfigure}[b]{0.475\textwidth}
      \includegraphics[width=\textwidth]{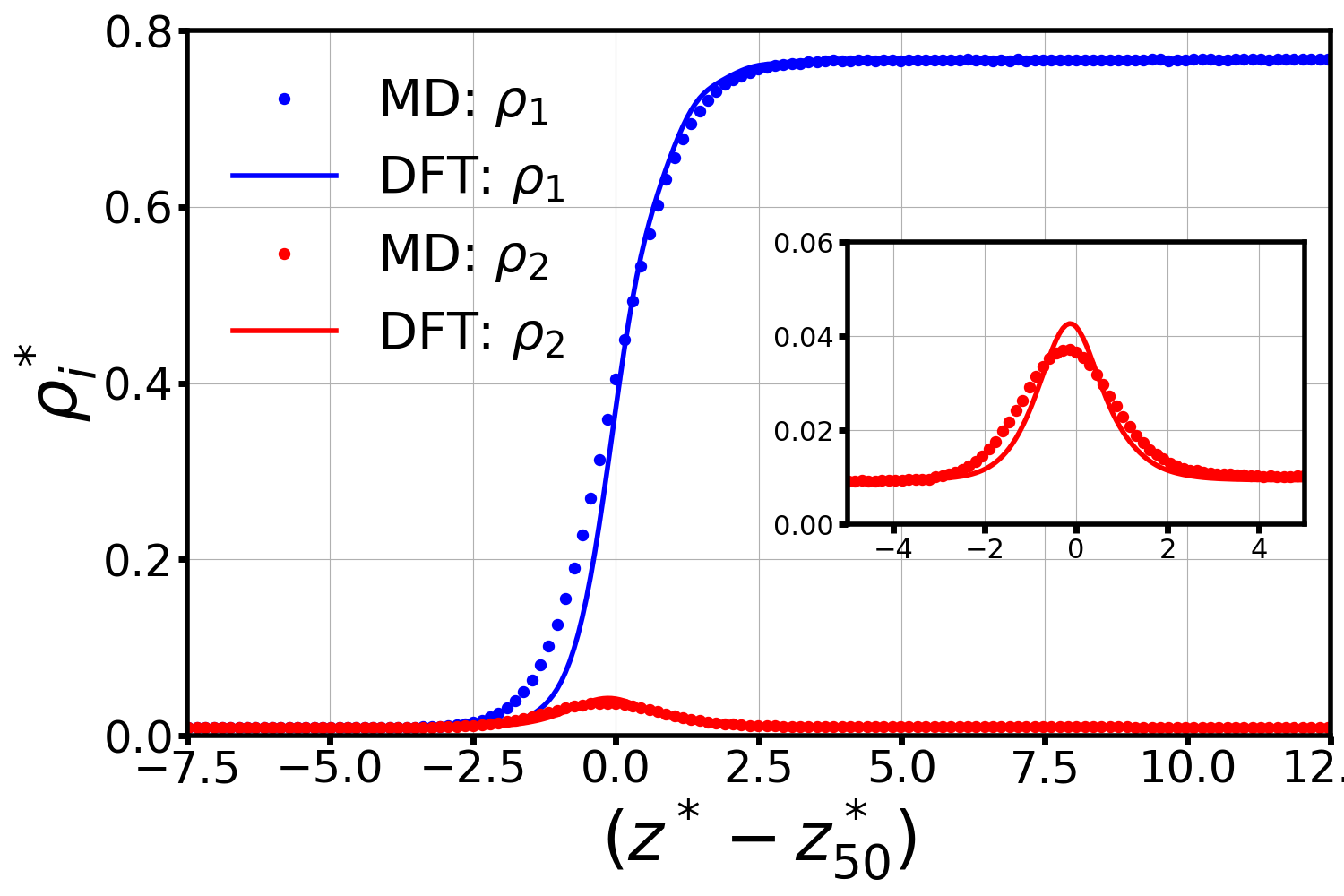}
      \caption{Mixture~A.}
      \label{fig:rho_eq_mix_A}
    \end{subfigure}
    \hfill
    \begin{subfigure}[b]{0.475\textwidth}
      \includegraphics[width=\textwidth]{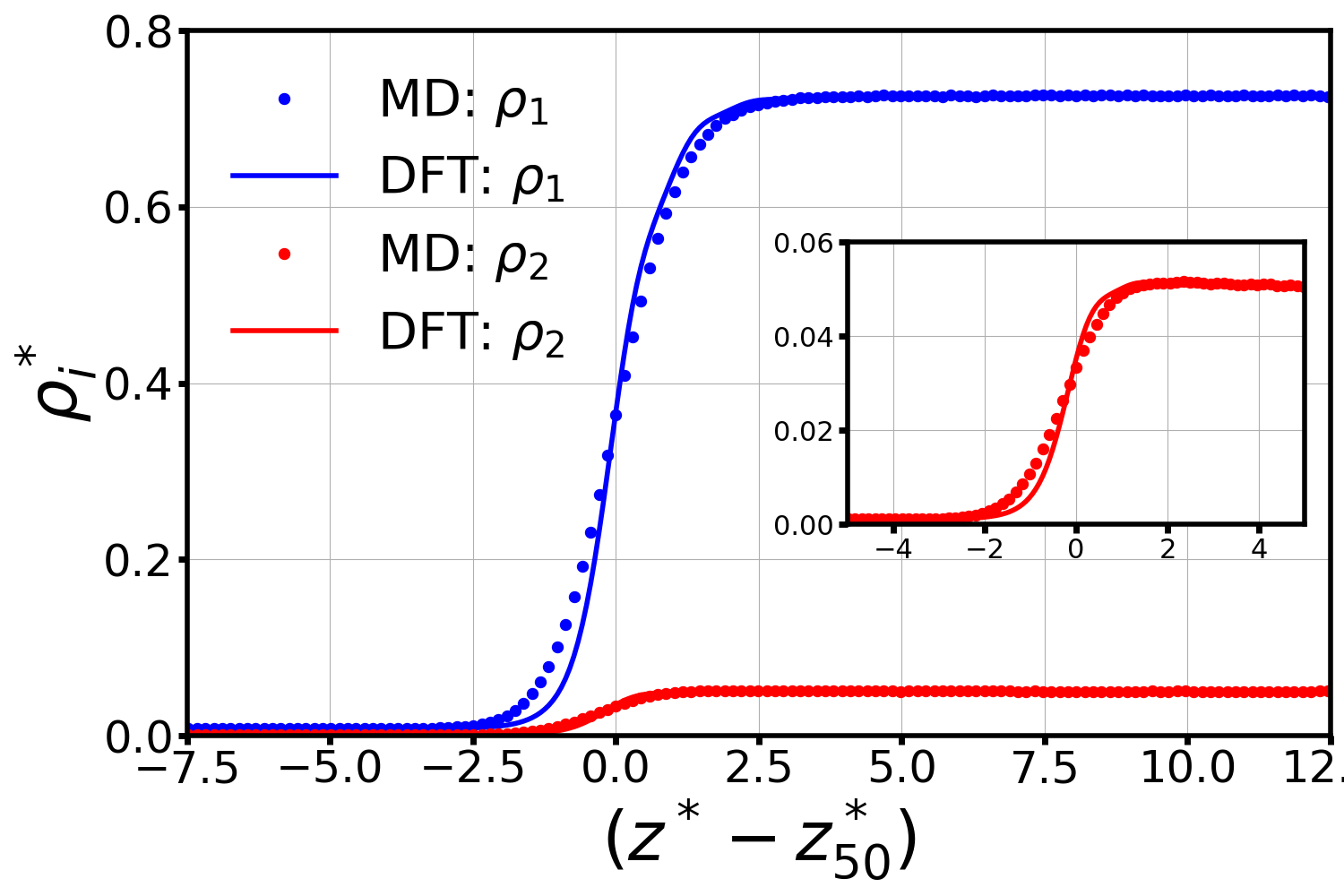}
      \caption{Mixture~B.}
      \label{fig:rho_eq_mix_B}
    \end{subfigure}
  
    \caption{Equilibrium density profiles of both components in both mixtures from DFT and MD at $T^*=0.715$. The inset magnifies the density of component 2.}
    \label{fig:rho_eq_mix}
  \end{figure}

  The equilibrium density profiles for both mixtures are shown in \cref{fig:rho_eq_mix}. Similar to the pure system (cf.\ \cref{fig:rho_eq_pure}), the density of the first component, $\rho_1^*$, increases monotonously from the vapor to the liquid density for both mixtures. However, the density of the second component, $\rho_2^*$, behaves differently in both mixtures. For mixture~A (see \cref{fig:rho_eq_mix_A}), $\rho_2^*$ exhibits a maximum at the interface; that is, the density at the interface is greater than in the bulk phases. This \emph{enrichment} is consistent with previous studies  \citep{stephan2019interfacial}. No enrichment is observed for mixture~B (see \cref{fig:rho_eq_mix_B}). The density of the second component, $\rho_2^*$, has a similar shape to that of the first component, $\rho_1^*$. 
  From \cref{fig:rho_eq_pure,fig:rho_eq_mix} we conclude that DFT accurately predicts the initial and final density profiles in the system. This is a prerequisite for meaningful comparisons of dynamic results in subsequent sections. 
  
  \subsection{Density and Flux in Measurement Volumes}

  \begin{figure*}
    \centering
    \begin{subfigure}[t]{0.475\textwidth}
      \caption{Vapor: density $\rho_2^*$ in measurement volume MV$_\text{vap}$.}
      \includegraphics[width=\textwidth,trim={0 0cm 0 0cm}, clip]{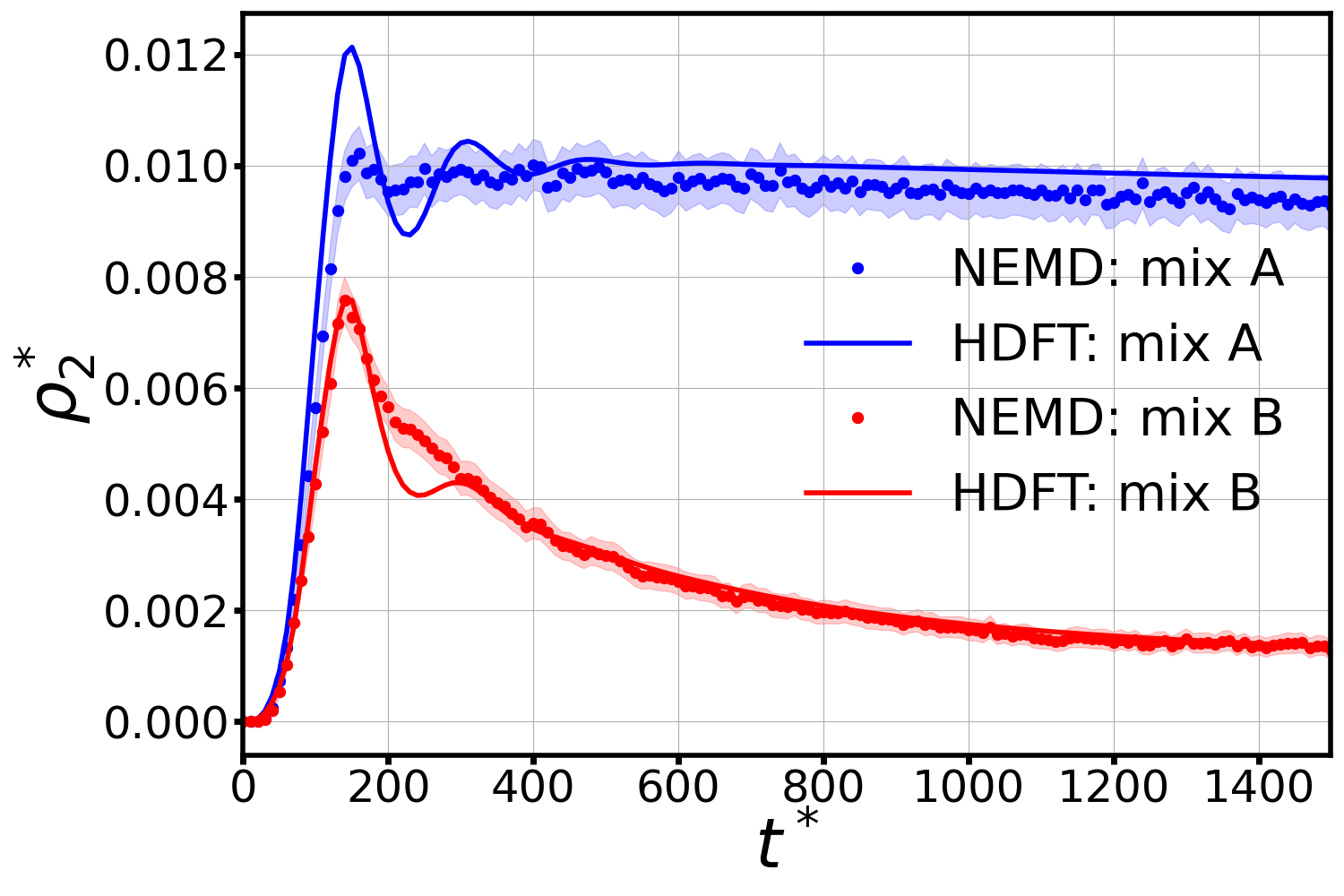}
      \label{fig:mv_rho_vap}
    \end{subfigure}
    \hfill
    \begin{subfigure}[t]{0.475\textwidth}
      \caption{Liquid: density $\rho_2^*$ in measurement volume MV$_\text{liq}$.}
      \includegraphics[width=\textwidth,trim={0 0cm 0 0cm}, clip]{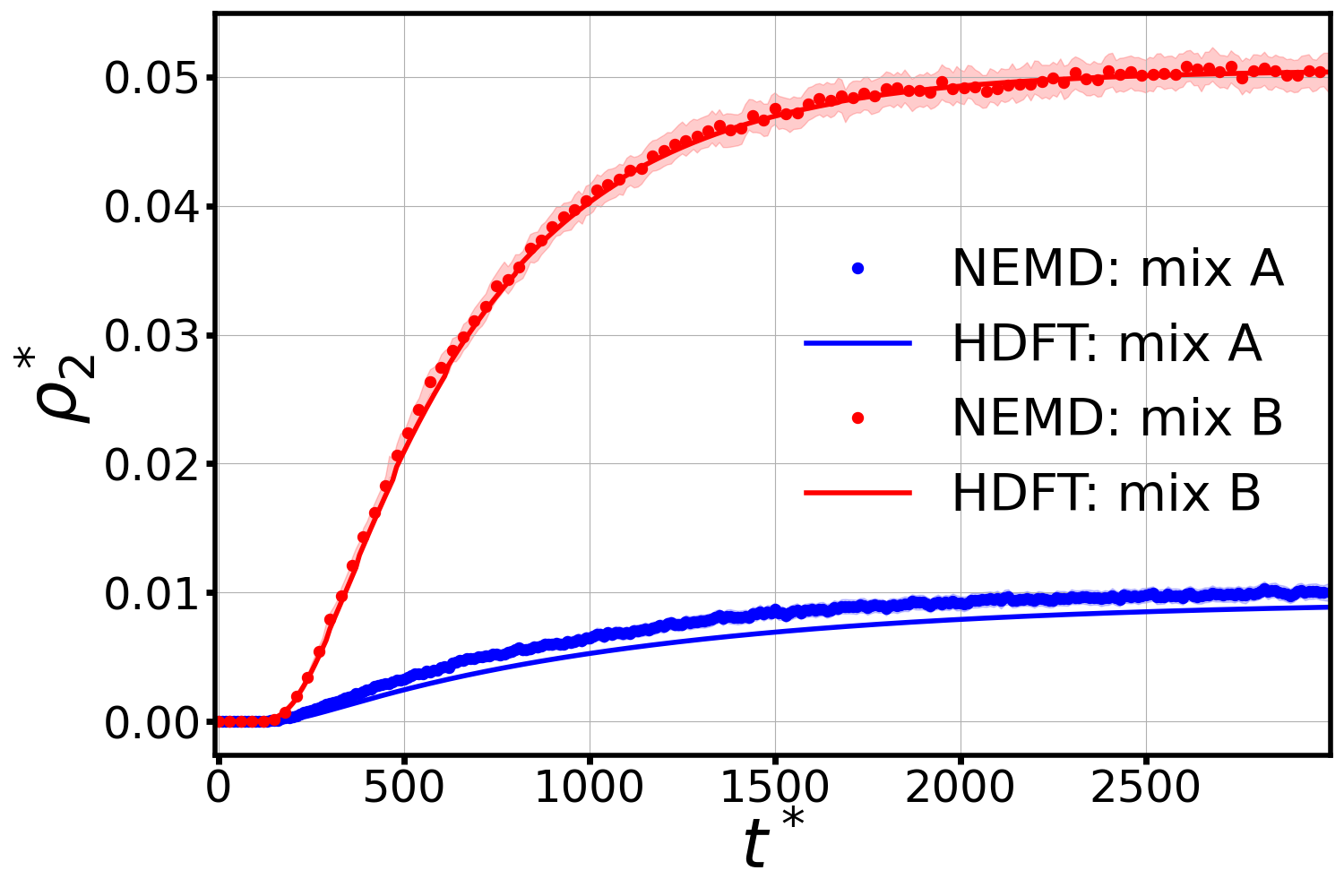}
      \label{fig:mv_rho_liq}
    \end{subfigure}
    \begin{subfigure}[t]{0.475\textwidth}
      \caption{Vapor: flux $j_2^*$ in measurement volume MV$_\text{vap}$.}
      \includegraphics[width=\textwidth,trim={0 0cm 0 0cm}, clip]{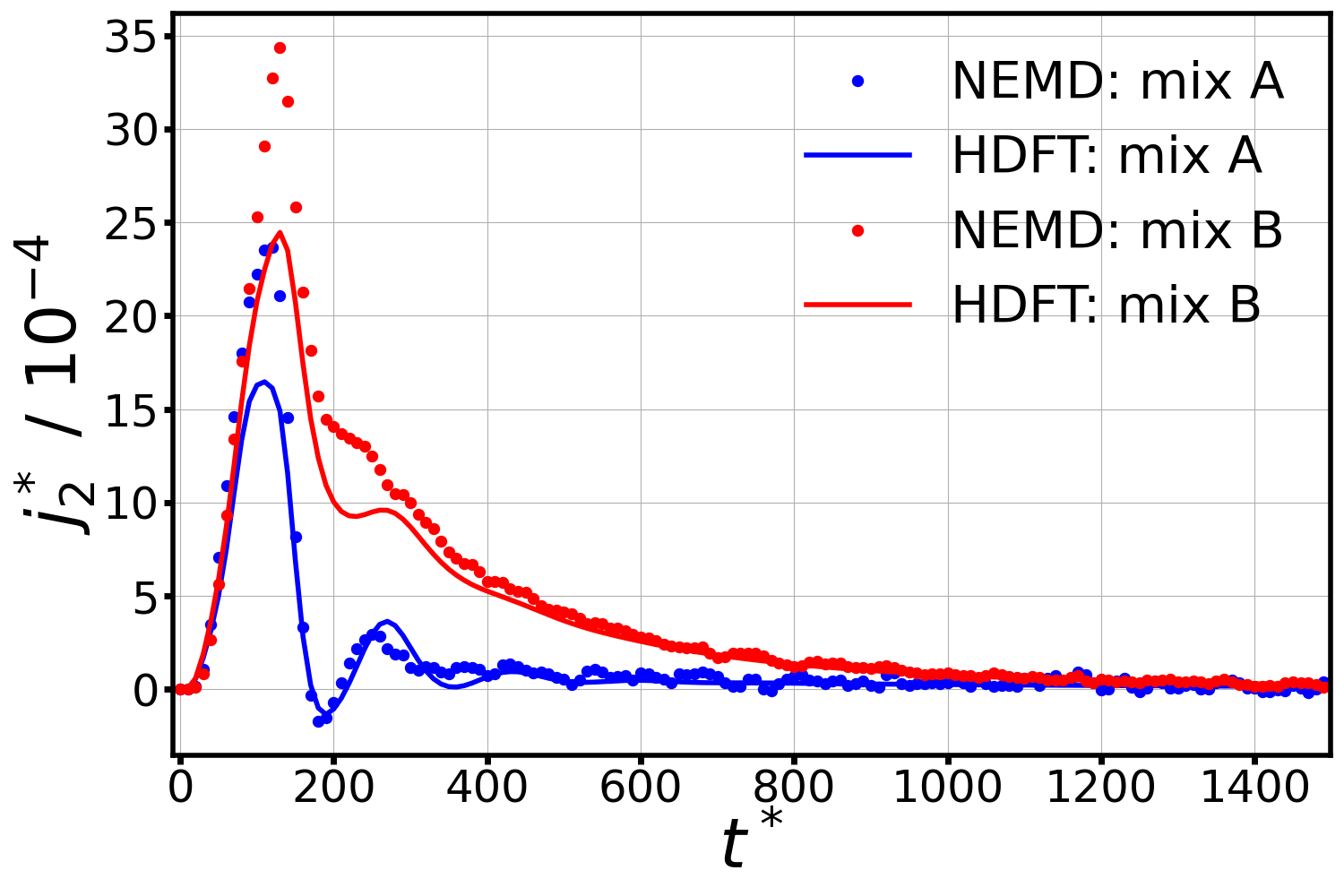}
      \label{fig:mv_flux_vap}
    \end{subfigure}
    \hfill
    \begin{subfigure}[t]{0.475\textwidth}
      \caption{Liquid: flux $j_2^*$ in measurement volume MV$_\text{liq}$.}
      \includegraphics[width=\textwidth,trim={0 0cm 0 0cm}, clip]{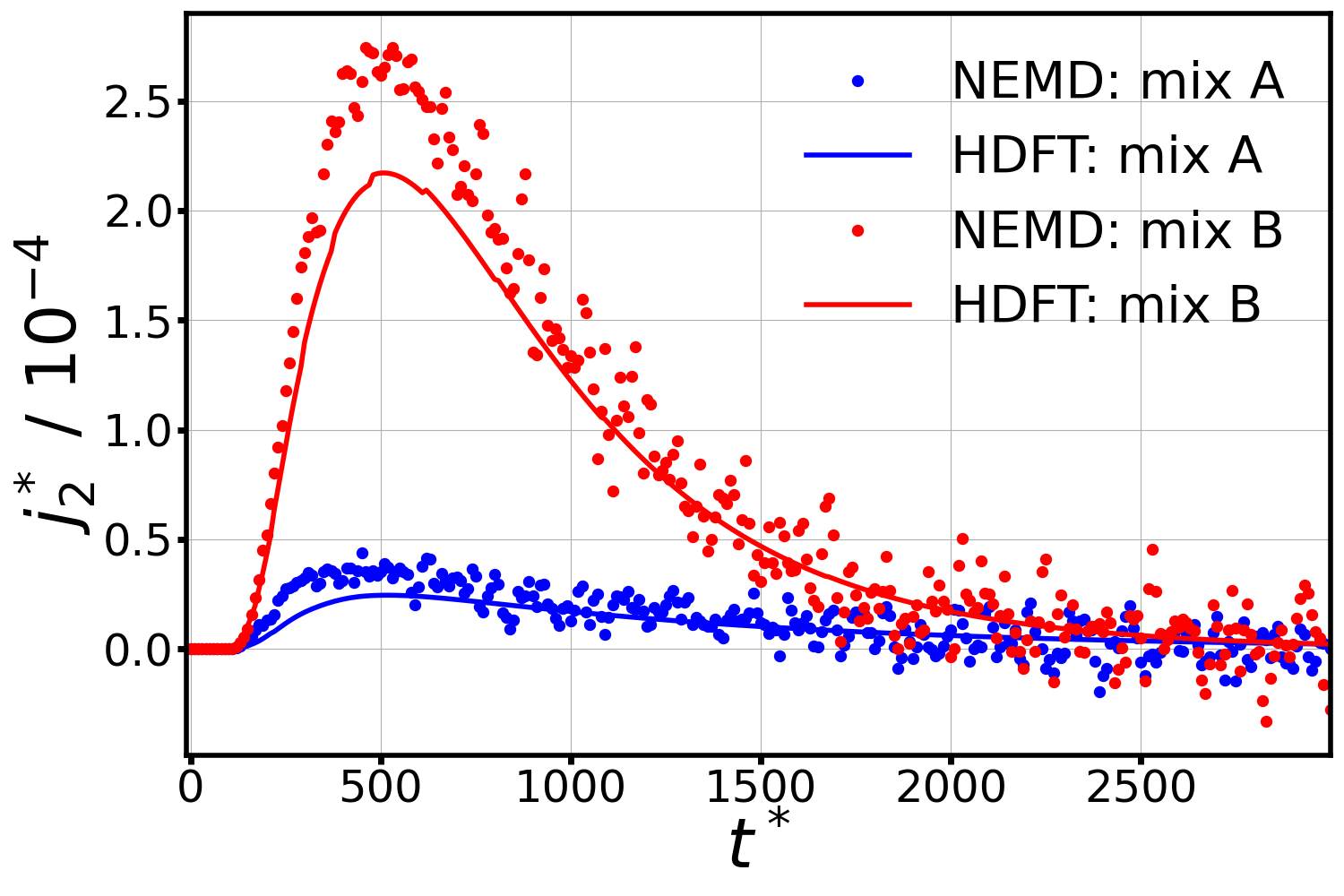}
      \label{fig:mv_flux_liq}
    \end{subfigure}
    \caption{Density $\rho_2^*$ and molar flux $j_2^*$ of component 2 over time $t^*$ in measurement volumes MV$_\mathrm{vap}$ and MV$_\mathrm{liq}$ for  mixture~A and mixture~B from hydrodynamic DFT and NEMD with the $95\%$ confidence interval (shaded regions) at $T^*=0.715$.}
    \label{fig:mv_all}
  \end{figure*}

  The density and flux response to the insertion, which occurs between $0\leq t^*\leq 100$, is determined for the measurement volumes and is depicted in \cref{fig:mv_all}. The density $\rho_2^*$ in the vapor phase (\cref{fig:mv_rho_vap}) from hydrodynamic DFT shows a rapid increase shortly after the insertion begins and reaches a maximum (around $t^*\approx 150$) for both mixtures. 
  For mixture~A, the maximum density is larger. 
  Mixture~B decreases more quickly after reaching its maximum.
  The temporary maximum in density is a dynamic phenomenon caused by the rapid insertion of component 2 molecules into the center of the vapor region. These molecules are transported toward the interface by a gradient in the chemical potential. 
  The higher maximum density and slower decay for mixture~A are consistent with the phase behavior of the mixtures. In mixture~A, mollecules of component 2 tend to remain at the interface and in the vapor phase more than in mixture~B. Thus, fewer molecules transfer from the vapor phase to the interface and the density in mixture~A remains close to the initial maximum. In contrast, the molecules of component 2 in mixture~B tend to enter the liquid phase, causing the initial density peak to decay rapidly. 
  Similar behavior is observed for both mixtures in the NEMD results, although the oscillations are less pronounced as compared to results from hydrodynamic DFT. 
  
  The behavior of density in MV$_\mathrm{vap}$ can be further analyzed based on the corresponding molecular flux (see \cref{fig:mv_flux_vap}), which is connected to density via component balances (\cref{eq:ComponentDDFT}).
  For both mixtures, an initial maximum in the flux is observed, followed by a decrease toward zero. For mixture~A, hydrodynamic DFT predicts strongly oscillatory behavior, which aligns well with the NEMD results. However, for mixture~B the oscillations are much less pronounced and barely visible in the NEMD results. 
  For mixture~A, the decay of the flux is much faster, even attaining negative values, as reported by \citet{schaefer2023mass}. These observations are consistent with the results for the density in MV$_\mathrm{vap}$.
  
  The oscillations in density and flux, as well as the negative flux values can be explained by the fact that molecules are `repelled' at the interface. This occurs because  a large number of molecules arrive at the interface, but they tend to avoid entering the liquid phase. Consequently, the flux is reduced and can even become negative, meaning the direction of the flux is reversed. Due to the symmetry of the system, this leads to oscillating behavior. Because the molecules of component 2 have a higher tendency to enter the liquid phase in mixture~B, these oscillations are much less pronounced compared to micture~A. 
  
  The density and flux in the  liquid phase measurement volume (MV$_\text{liq}$) are shown in \cref{fig:mv_rho_liq} and \cref{fig:mv_flux_liq}, respectively. Compared to the vapor phase, the density increases later, due to the time required for mass transfer through the vapor-liquid interface. The density increases continuously and approaches a constant (equilibrium) value toward the end of the simulation ($t^*\approx 3000$). Consistent with the equilibrium densities shown in \cref{fig:rho_eq_mix}, the density at the end of the simulation is much larger in mixture~B than in mixture~A. For mixture~A, the density values differ between hydrodynamic DFT and NEMD at the end of the simulation. This difference stems from discrepancies between the PeTS equation of state and the MD simulations of the LJTS fluid, i.e.\ different densities in the bulk phases.  
  The flux shows a maximum for both mixtures and monotonically decreases toward zero (see \cref{fig:mv_flux_liq}). 
  Overall, these results demonstrate that the densities and fluxes from hydrodynamic DFT in both phases and mixtures are in good agreement  with results from NEMD, which suggests that mass transfer across the vapor-liquid interface is accurately predicted. 
  
  \begin{figure}
    \centering
    \begin{subfigure}[b]{0.475\textwidth}
      \includegraphics[width=\textwidth,trim={0 0cm 0 0cm}, clip]{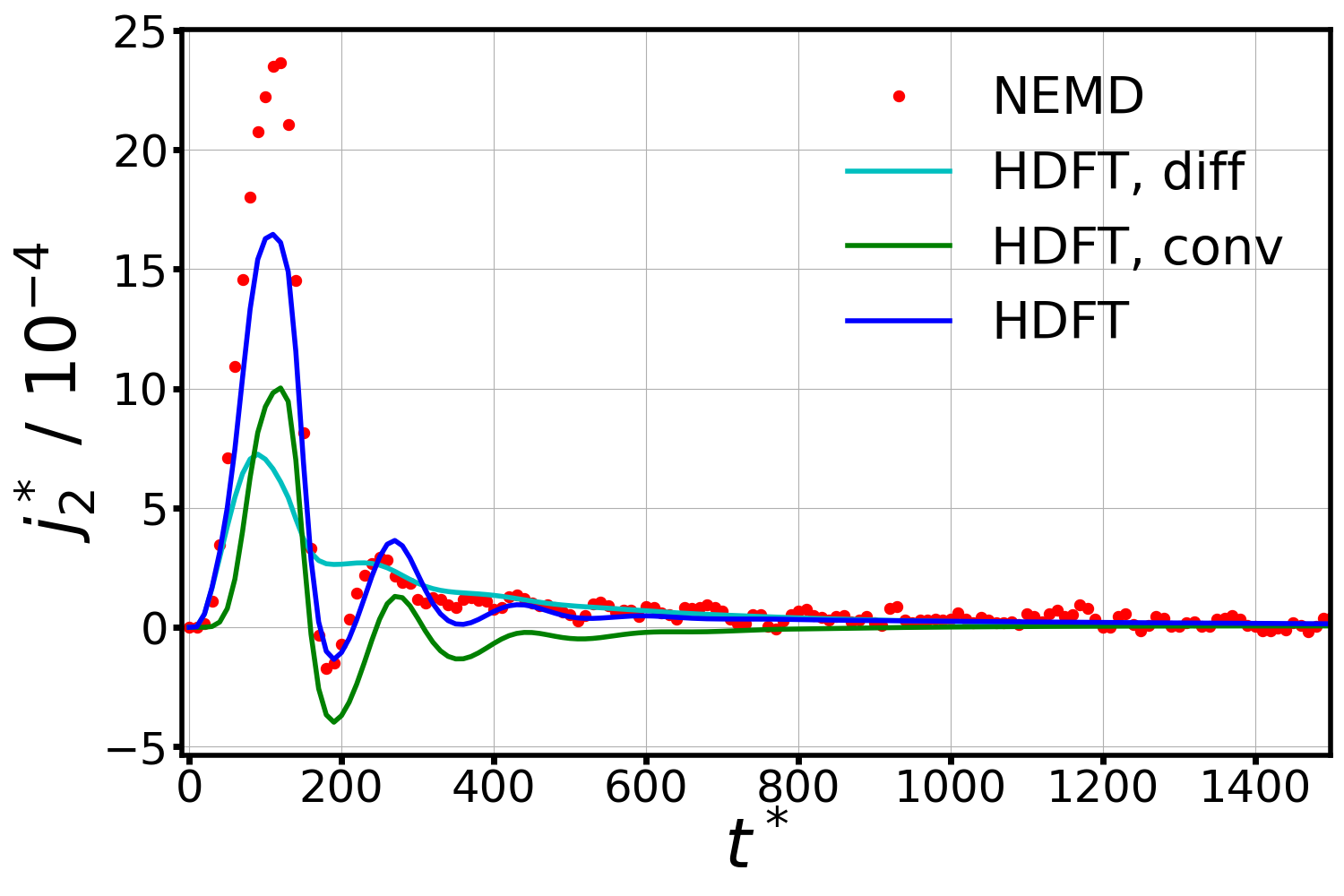}
      \caption{Vapor: flux $j_2^*$ in measurement volume MV$_\text{vap}$.}
      \label{fig:flux_conv_diff_vap}
    \end{subfigure}
    \hfill
    \begin{subfigure}[b]{0.475\textwidth}
      \includegraphics[width=\textwidth,trim={0 0cm 0 0cm}, clip]{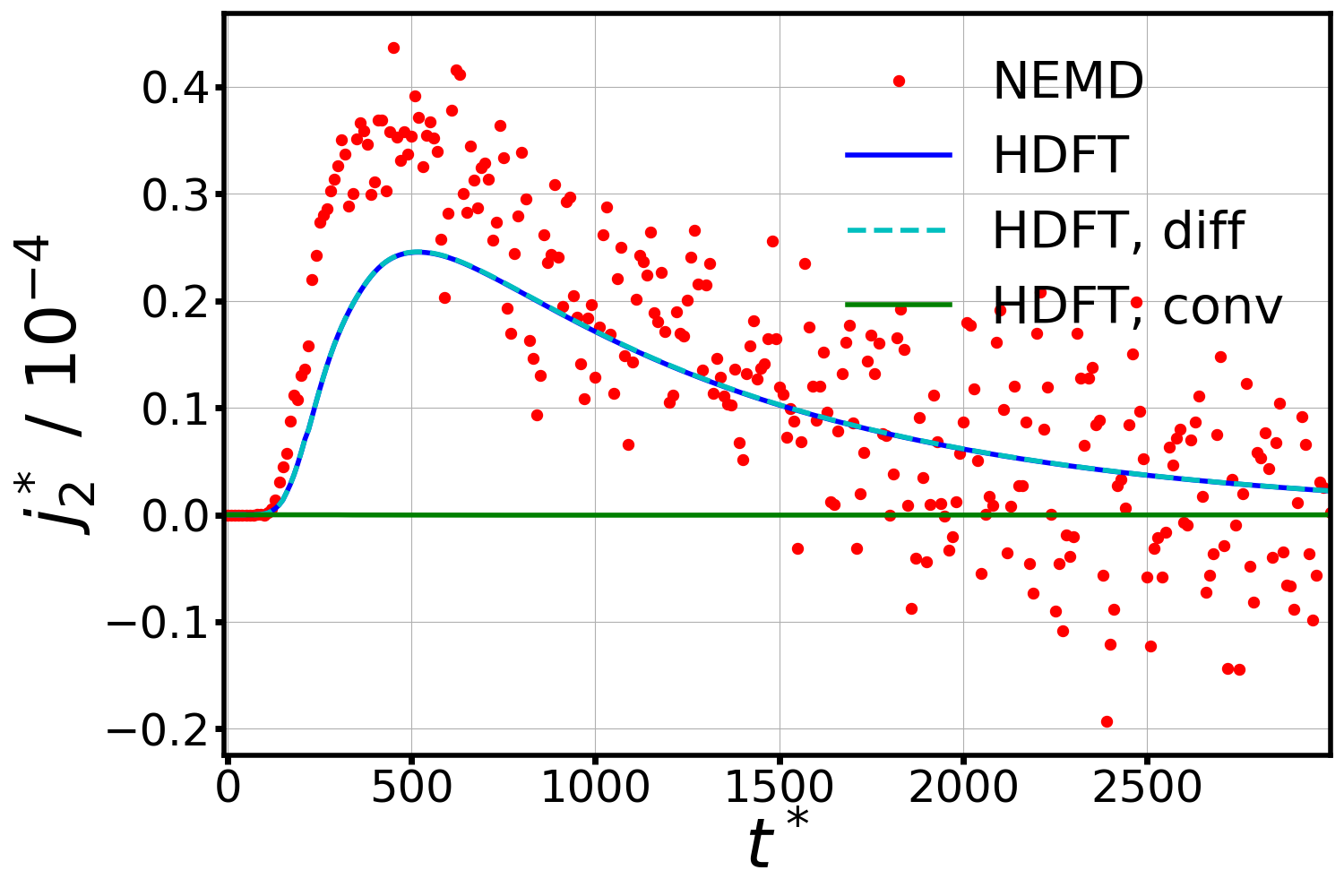}
      \caption{Liquid: flux $j_2^*$ in measurement volume MV$_\text{liq}$.}
      \label{fig:flux_conv_diff_liq}
    \end{subfigure}
    \caption{Flux profiles $j_2^*$ for mixture~A from hydrodynamic DFT and NEMD, where the total flux from hydrodynamic DFT is separated into convective and diffusive contributions at $T^*=0.715$.}
    \label{fig:flux_conv_diff}
  \end{figure}
  
  Hydrodynamic DFT provides additional insight into transport mechanisms by enabling the separation of convective and diffusive flux contributions, as illustrated in \cref{fig:flux_conv_diff}. Achieving such a separation with a reasonable signal-to-noise ratio is challenging in NEMD simulations.
  In MV$_\mathrm{vap}$ (see \cref{fig:flux_conv_diff_vap}), the diffusive flux is always positive, i.e., directed toward the interface, and exhibits only minor oscillations. In contrast, the convective flux shows pronounced oscillations and becomes negative for significant periods of time. These results support the hypothesis that molecules are repelled at the interface, resulting temporarily in a negative convective flux (i.e.\ a flux directed away from the interface). Because the diffusive flux remains positive, the total flux (the sum of the diffusive and convective contributions) is only briefly negative, which occurs exclusively for mixture~A.
  In MV$_\mathrm{liq}$ (see \cref{fig:flux_conv_diff_liq}) the convective flux is close to zero, and the total flux  equals the diffusive flux. This suggests that transport in the liquid phase is solely determined by diffusion.

  \subsection{Density and Flux Profiles}
  
  \begin{figure*}
    \centering
    \begin{subfigure}[t]{0.475\textwidth}
      \caption{Mixture~A: density $\rho_2^*$ profile.}
      \includegraphics[width=\textwidth,trim={0 0cm 0cm 0}, clip]{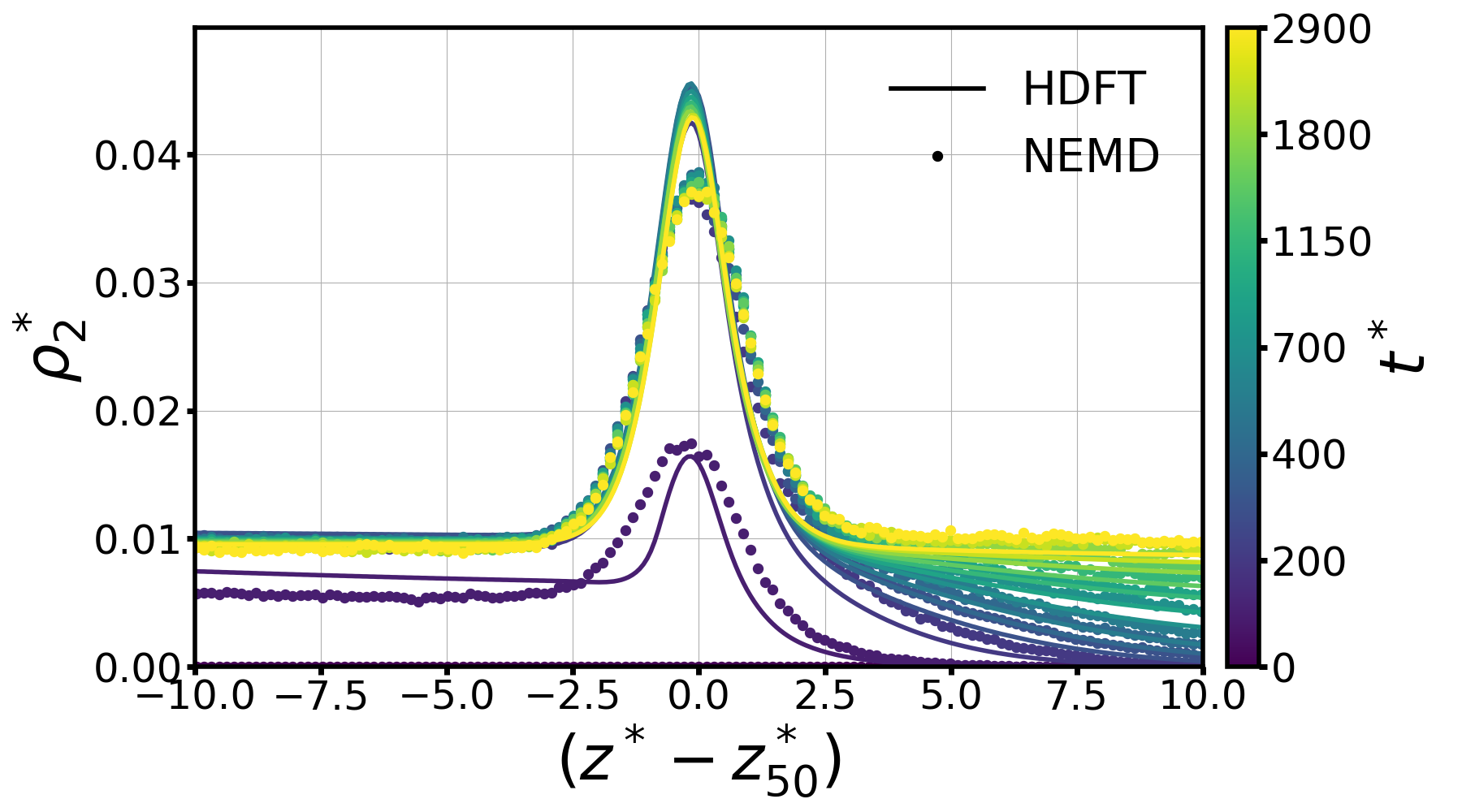}
      \label{fig:profiles_rho_A}
    \end{subfigure}
    \hfill
    \begin{subfigure}[t]{0.475\textwidth}
      \caption{Mixture~B: density $\rho_2^*$ profile.}
      \includegraphics[width=\textwidth]{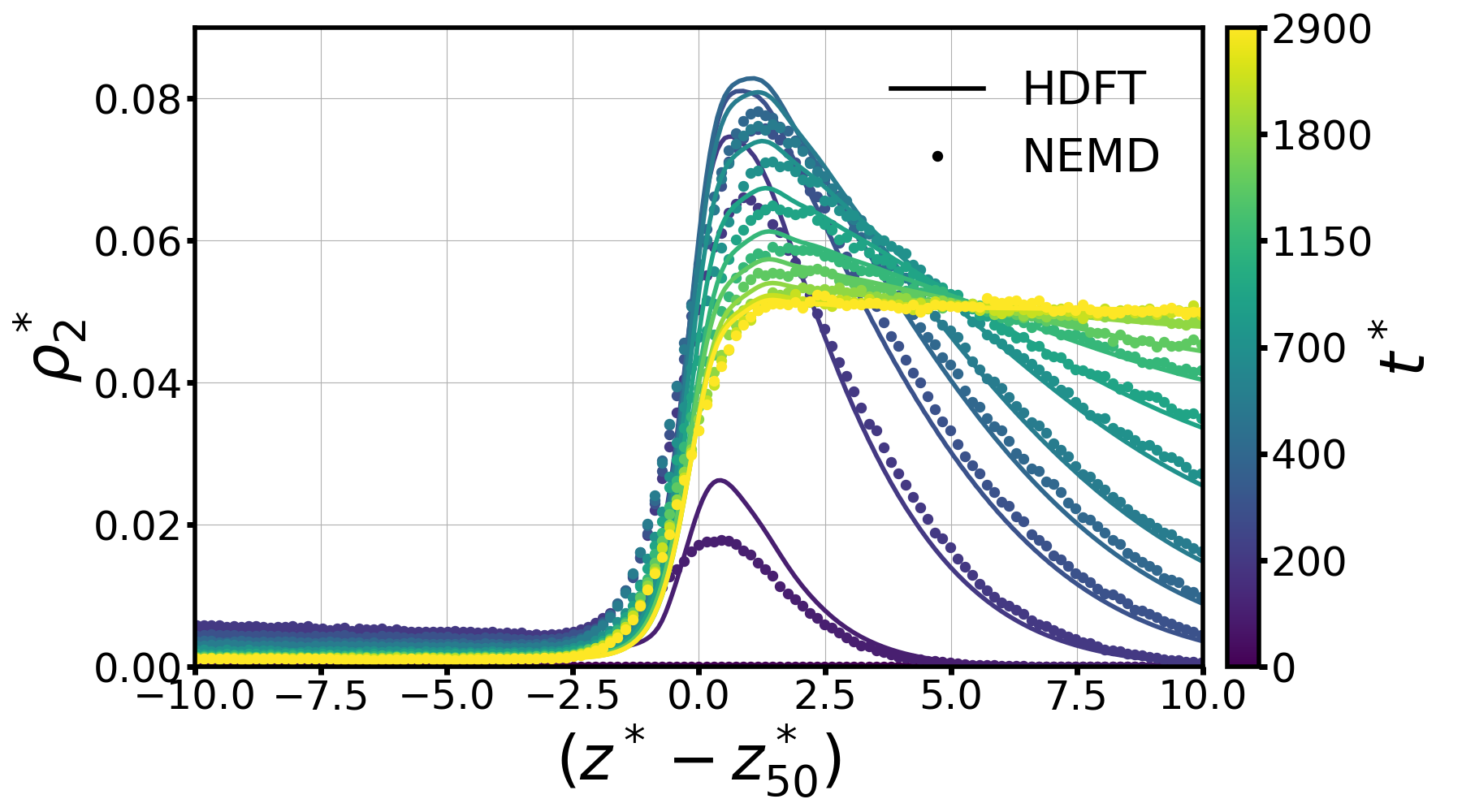}
      \label{fig:profiles_rho_B}
    \end{subfigure}
    \begin{subfigure}[t]{0.475\textwidth}
      \caption{Mixture~A: flux $j_2^*$ profile.}
      \includegraphics[width=\textwidth,trim={0 0cm 0 0cm}, clip]{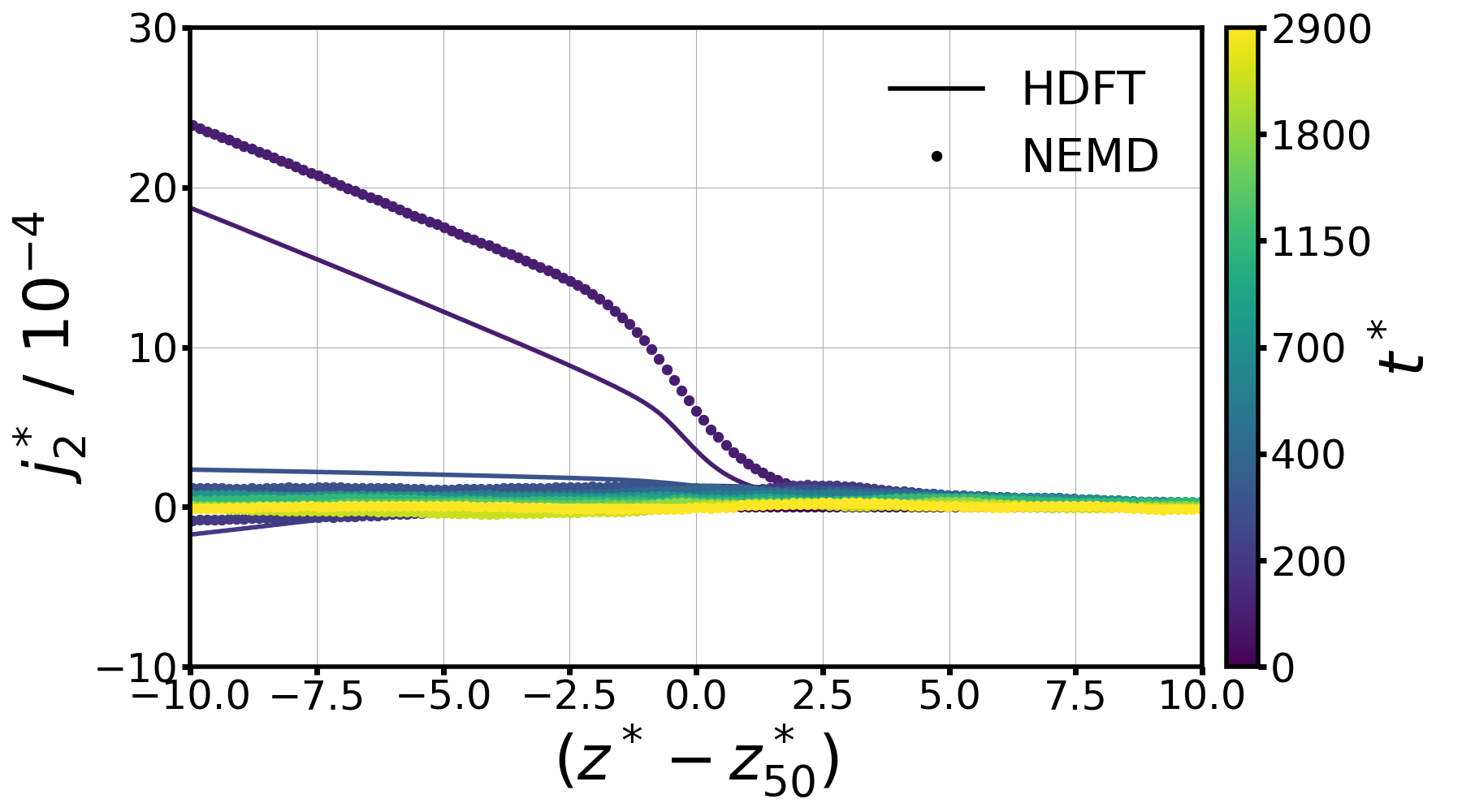}
      \label{fig:profiles_flux_A}
    \end{subfigure}
    \hfill
    \begin{subfigure}[t]{0.475\textwidth}
      \caption{Mixture~B: flux $j_2^*$ profile.}
      \includegraphics[width=\textwidth,trim={0 0cm 0 0cm}, clip]{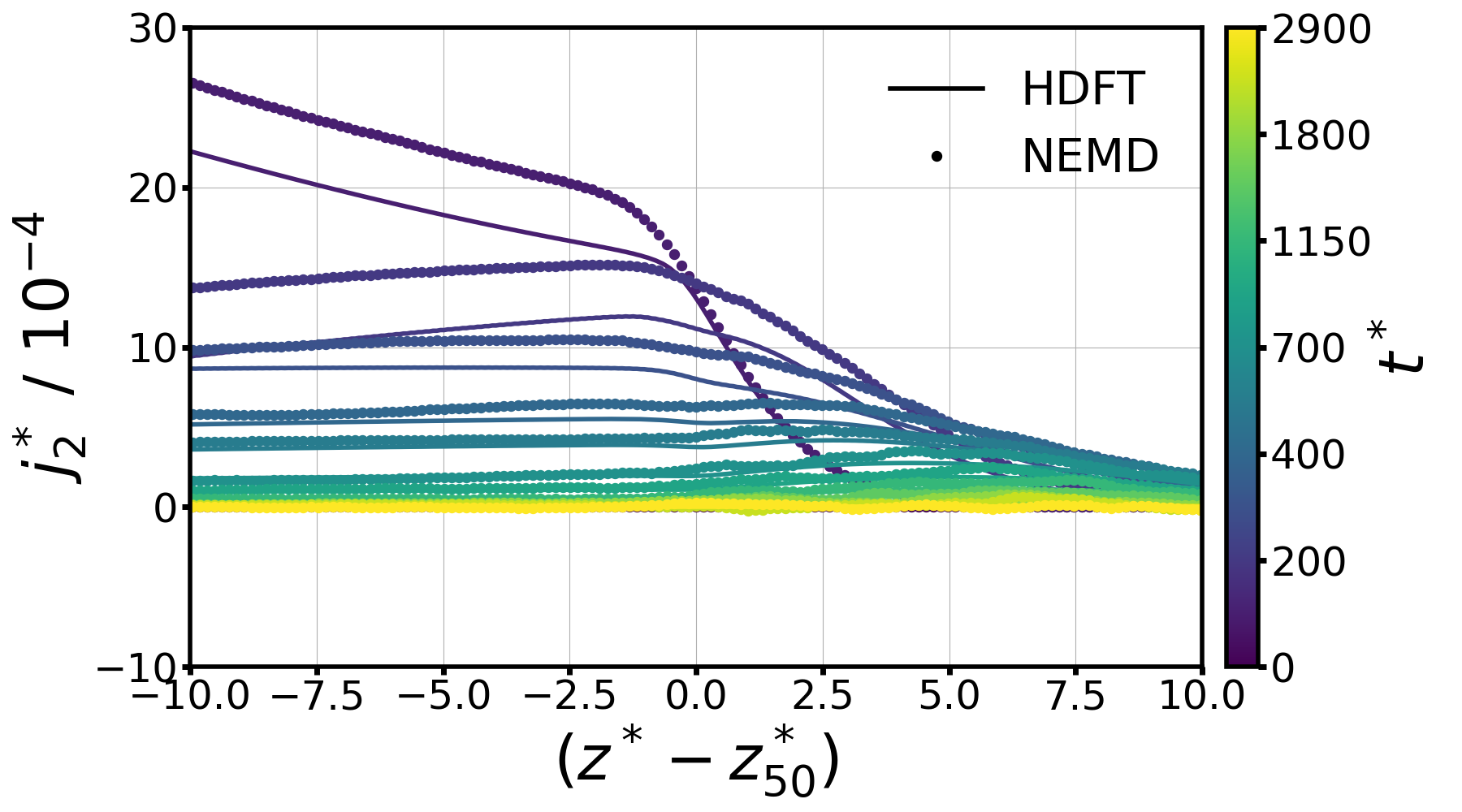}
      \label{fig:profiles_flux_B}
    \end{subfigure}
  
    \caption{Density $\rho_2^*$ and flux $j_2^*$ profiles across the interface during relaxation from hydrodynamic DFT and NEMD at $T^*=0.715$.}
    \label{fig:profiles}
  \end{figure*}
  
  In addition to the results in the two measurement volumes, we aim to assess whether hydrodynamic DFT can predict the behavior across the entire interface. \Cref{fig:profiles} shows predictions of  hydrodynamic DFT for the density $\rho_2^*$ and flux $j_2^*$ profiles across the interface compared to NEMD results at several time steps. Lighter colors indicate later times in the simulation.
  For mixture~A (see \cref{fig:profiles_rho_A}), the density $\rho_2^*$ at the interface increases and forms a peak similar to the equilibrium density profile (cf.\ \cref{fig:rho_eq_mix}). The density in the liquid phase steadily increases until the equilibrium density is reached. 
  The corresponding flux profile (\cref{fig:profiles_flux_A}) shows that the flux is initially large in the vapor phase and within the interface. As was previously observed for MV$_\mathrm{vap}$ (cf. \cref{fig:mv_flux_vap}), the flux oscillates and becomes negative before quickly approaching zero. 
  
  For mixture~B (see \cref{fig:profiles_rho_B}), the density $\rho_2^*$  within the interface also forms a peak, which is in contrast to the  equilibrium density profile, which shows no such peak. 
  Toward the end of the simulation  the density peak disappears and the liquid phase density increases, resulting in the equilibrium density profile. This temporary density peak can be considered \emph{temporary enrichment}.
  The flux across the interface for mixture~B (see \cref{fig:profiles_flux_B}) is large at the beginning of the simulation and decreases more slowly than for mixture~A.  
  Interestingly, the flux gradient reverses and becomes positive (initially negative) for some intermediate profiles.
   
  The density profiles predicted by hydrodynamic DFT are in good agreement with the NEMD results, and all phenomena, such as temporary enrichment, are accurately reproduced. The flux profiles exhibit larger deviations, but are still captured sufficiently well. Note that the NEMD flux profiles are obtained by numerically differentiating the density and solving the component balances within each bin, calculations that are subject to statistical uncertainties.

  \subsection{Influence of Temperature}
  
  \begin{figure*}
    \centering
    \begin{subfigure}[t]{0.475\textwidth}
      \caption{Vapor: density $\rho_2^*$ in MV$_\text{vap}$ for mixture~A.}
      \label{fig:mv_T_rho_vap_mixA}
      \includegraphics[width=\textwidth,trim={0 0cm 0 0cm}, clip]{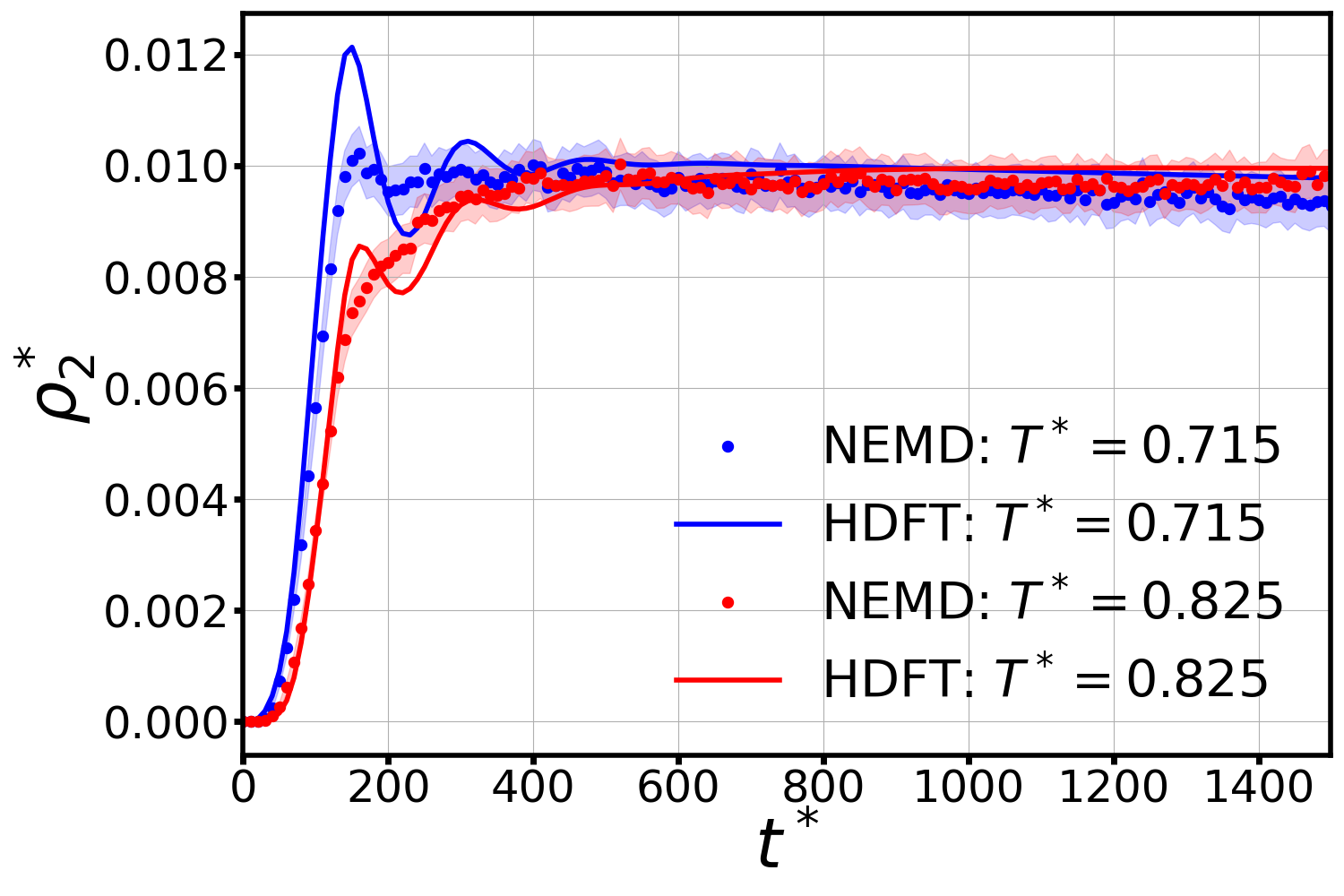}
    \end{subfigure}
    \hfill
    \begin{subfigure}[t]{0.475\textwidth}
      \caption{Liquid: density $\rho_2^*$ in MV$_\text{liq}$ for mixture~A.}
      \label{fig:mv_T_rho_liq_mixA}
      \includegraphics[width=\textwidth,trim={0 0cm 0 0cm}, clip]{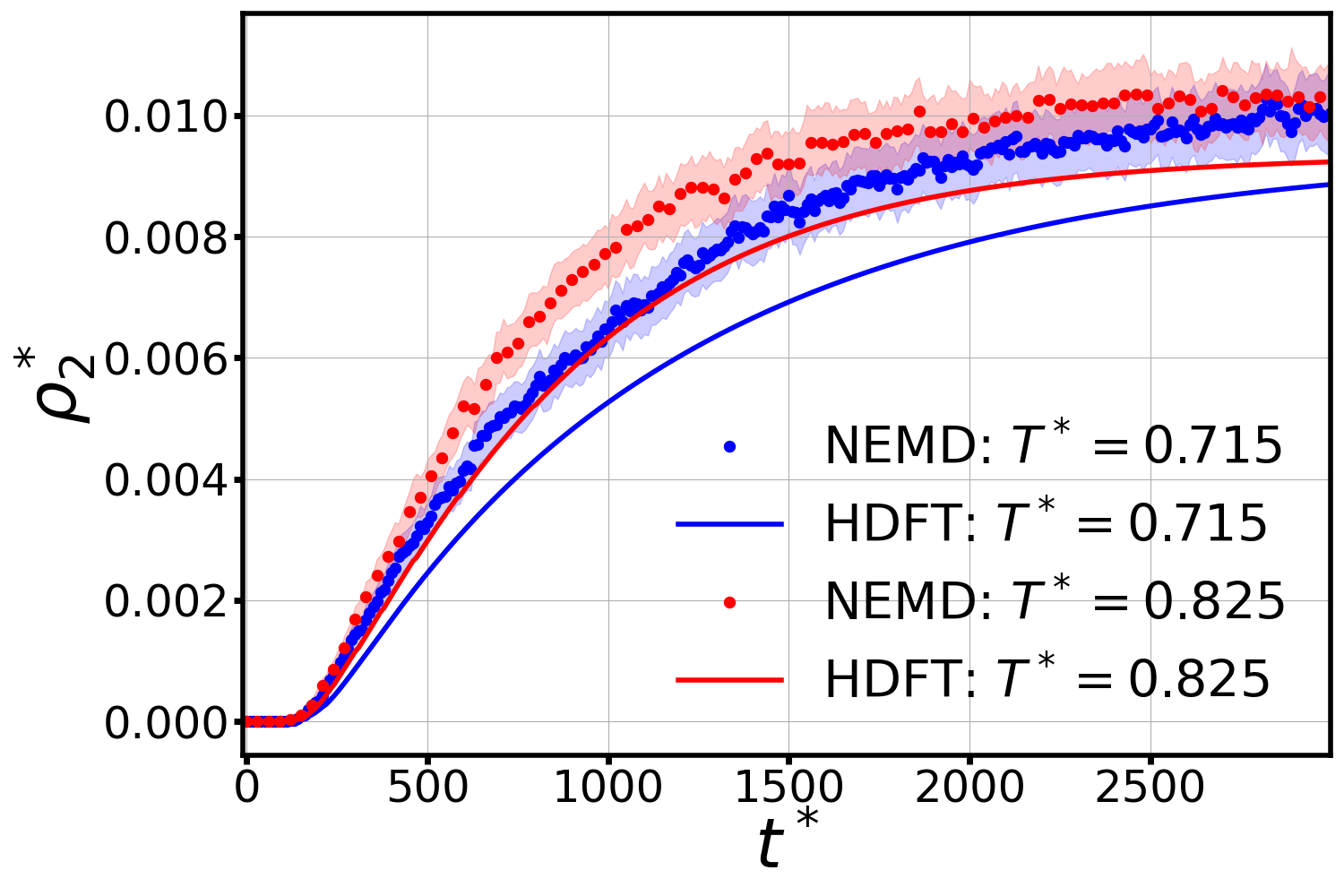}
    \end{subfigure}
    \begin{subfigure}[t]{0.475\textwidth}
      \caption{Vapor: density $\rho_2^*$ in MV$_\text{vap}$ for mixture~B.}
      \label{fig:mv_T_rho_vap_mixB}
      \includegraphics[width=\textwidth,trim={0 0cm 0 0cm}, clip]{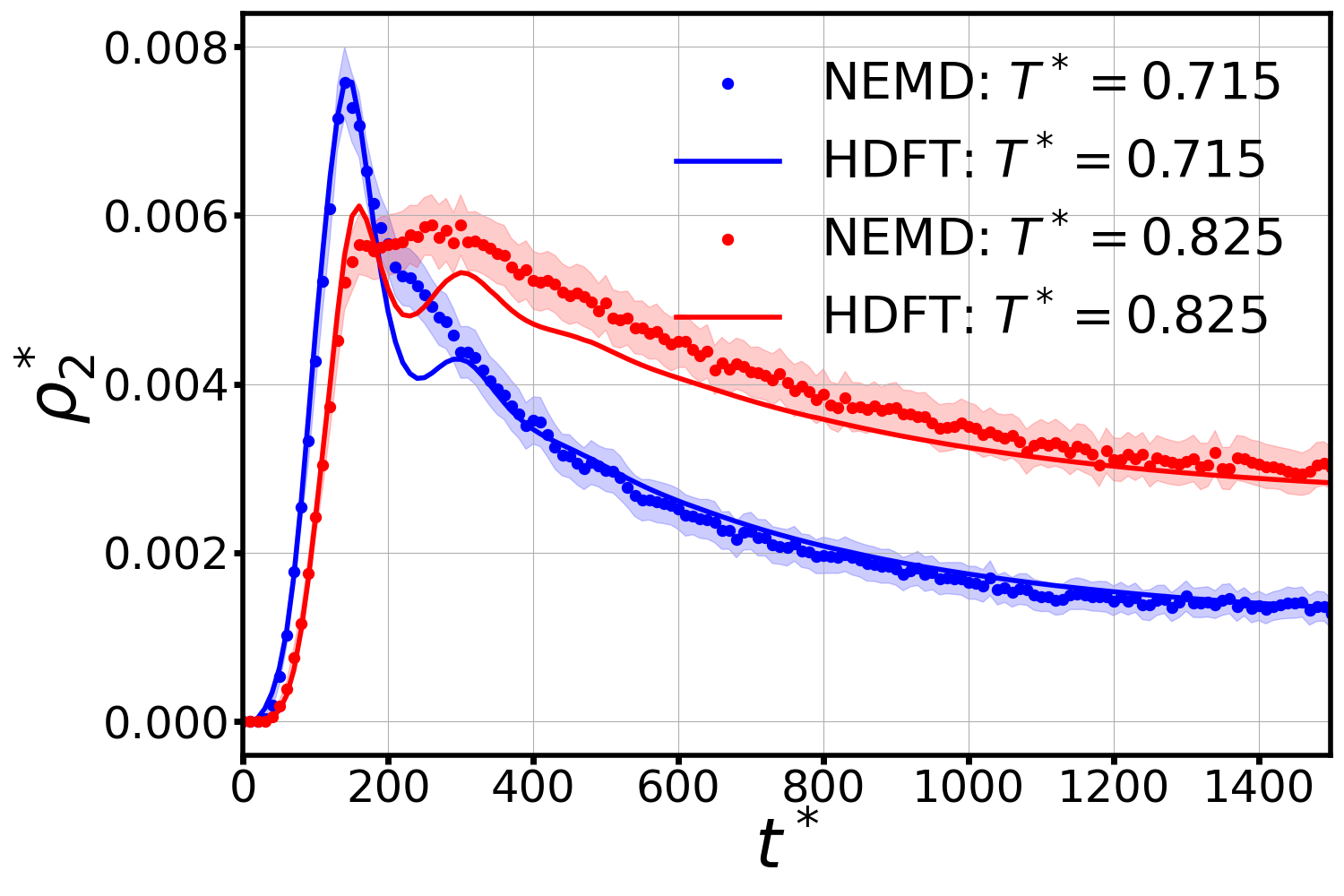}
    \end{subfigure}
    \hfill
    \begin{subfigure}[t]{0.475\textwidth}
      \caption{Liquid: density $\rho_2^*$ in MV$_\text{liq}$ for mixture~B.}
      \label{fig:mv_T_rho_liq_mixB}
      \includegraphics[width=\textwidth,trim={0 0cm 0 0cm}, clip]{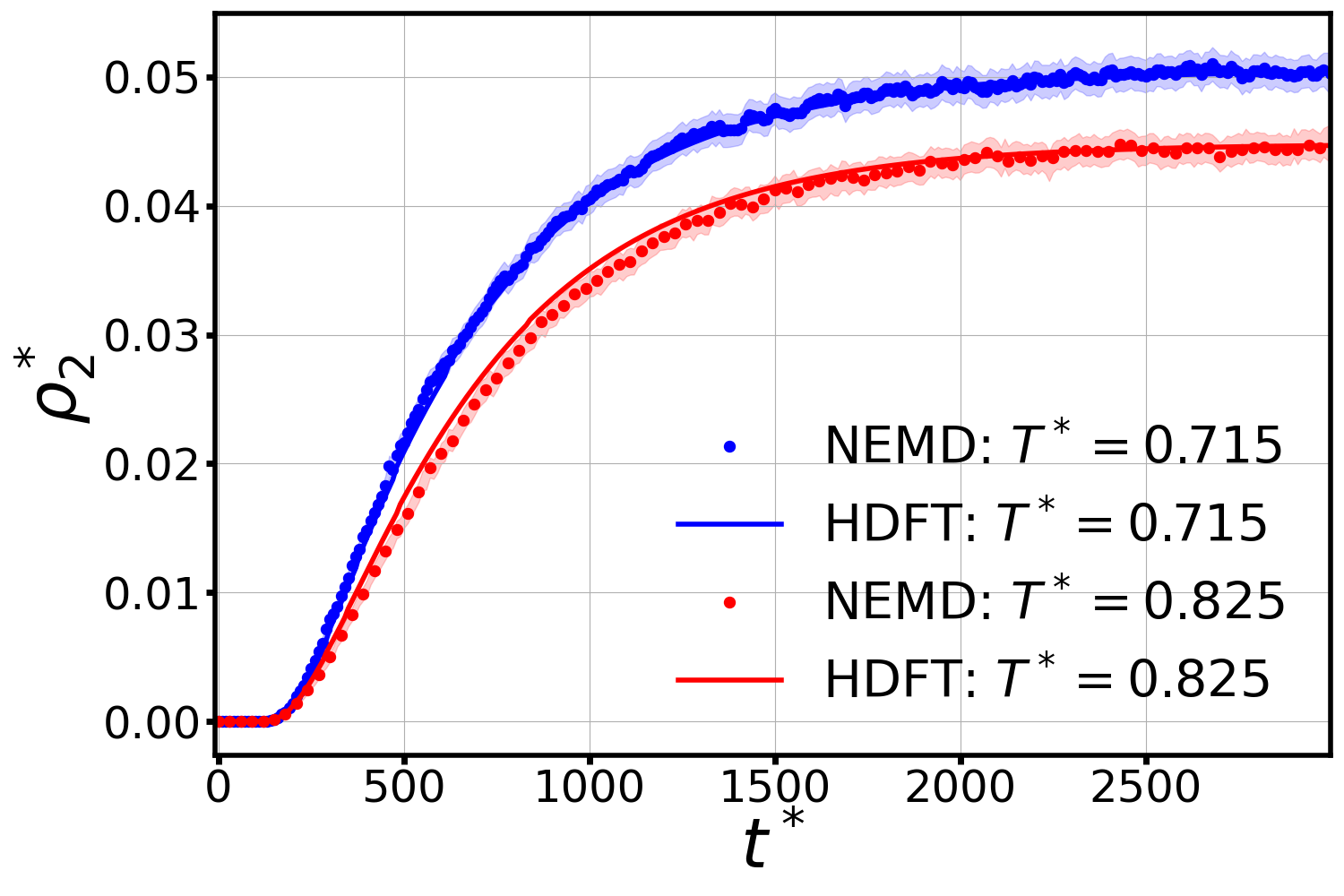}
    \end{subfigure}
    \caption{Density $\rho_2^*$  of component 2 over time $t^*$ in measurement volumes MV$_\mathrm{vap}$ and MV$_\mathrm{liq}$ for  mixture~A and mixture~B for $T^*=0.715$ and $T^*=0.825$ from hydrodynamic DFT and NEMD with the $95\%$ confidence interval (shaded region).}
    \label{fig:mv_T_all}
  \end{figure*}
  All simulations in this work are isothermal, and results have  thus far been presented for $T^* = 0.715$. \Cref{fig:mv_T_all} compares the densities in the two measurement volumes for both mixtures at an elevated temperature of $T^* = 0.825$ with those at $T^* = 0.715$.
  Increasing the temperature affects the first density peak in MV$_\mathrm{vap}$ for both mixtures. For mixture~A (see \cref{fig:mv_T_rho_vap_mixA}), the density $\rho_2^*$ does not show a peak and  approaches a value similar to that at $T^*=0.715$ toward the end of the simulation. For mixture~B (see \cref{fig:mv_T_rho_vap_mixB}), the density $\rho_2^*$  shows a lower initial peak and then decreases more slowly, but to a higher value, than at $T^*=0.715$. 
  In the liquid phase (MV$_\mathrm{liq}$), the density $\rho_2^*$ in mixture~A (see \cref{fig:mv_T_rho_liq_mixA}) increases slightly faster at higher temperatures but converges to a similar value at both temperatures. However, in mixture~B (see \cref{fig:mv_T_rho_liq_mixB}), the density $\rho_2^*$ increases faster and converges toward a larger value at a lower temperature. 
  
  This behavior is governed by two aspects.  
  First, the equilibrium density of component~2 in the mixture determines the asymptotic density value observed at the end of the simulation. For mixture~B, the equilibrium density exhibits a temperature dependence, whereas this dependence is negligible for mixture~A (see the Supporting Information). 
  Second, interfacial enrichment is more pronounced at lower temperatures. It is permanent for mixture~A and  temporary for mixture~B.   
  Interestingly, mixture~A exhibits similar equilibrium densities  at both temperatures, which cannot explain the temperature dependence of the initial density peak in MV$_\mathrm{vap}$. This suggests that interfacial enrichment significantly impacts mass transfer across the interface \citep{schaefer2023mass}. 
  
  The results from hydrodynamic DFT differ from NEMD results for the density $\rho_2^*$ in the liquid phase of mixture~A. As mentioned above, the equilibrium densities obtained from the PeTS equation of state differ from MD results, and this difference extends to the dynamic case. Additionally, the density in the vapor phase obtained from hydrodynamic DFT exhibits larger oscillations than NEMD results, as is the case at lower temperatures. Nevertheless, there is good agreement between the hydrodynamic DFT and NEMD results for both temperatures and both mixtures. The hydrodynamic DFT model correctly captures the effect of temperature on the interplay between equilibrium density, interfacial enrichment, and the density's dynamic behavior. We emphasize that the model does not include any temperature-dependent or mass-transfer-related adjustable parameters.

  \section{Conclusion} \label{sec:conclusion}
  
  In this work we have investigated whether hydrodynamic DFT is able to predict mass transfer across vapor-liquid interfaces by inserting molecules of a second component into an initially pure system. 
  The hydrodynamic DFT approach is combined with a generalized version of entropy scaling to determine local values for the transport properties, namely the viscosity as well as the self-diffusion and Maxwell-Stefan diffusion coefficients. The influence of the vapor-liquid interface on the dynamics of the system is captured by the DFT term in the momentum balance. 
  The predictive capabilities of the hydrodynamic DFT model are assessed by comparing results for the density and molecular fluxes to NEMD simulations, both in measurement volumes close to the interface and across the interface.
  
  This investigation reveals good quantitative agreement between the hydrodynamic DFT and NEMD results for density and molecular fluxes throughout the system, especially across the interface. Three key results were obtained:
  First, mass transfer and related microscopic phenomena, such as temporary enrichment or repulsion of molecules at the interface, can be accurately predicted using hydrodynamic DFT. The diffusive and convective contributions to mass transfer can be unambiguously separated.  
  Second, the generalized entropy scaling approach for diffusion coefficients and the Maxwell-Stefan approach with a generalized driving force correctly describe diffusion in the individual phases and across the interface. 
  Third, hydrodynamic DFT is a versatile method that connects equilibrium phase behavior, molecular interactions, and mass transfer for different types of mixtures while accounting for varying (constant) temperatures.  
  Overall, hydrodynamic DFT provides a predictive framework for mass transfer across vapor–liquid interfaces and consistently reduces to the Navier–Stokes equations far from interfaces. This makes it well-suited for studying diffusion and interfacial mass transfer across multiple scales. Hydrodynamic DFT has direct applications to technical processes, such as absorption and distillation, as well as to multiphase transport in porous media.

\section*{Funding}
Funded by the Deutsche Forschungsgemeinschaft (DFG, German Research Foundation) – Project Number 327154368 – SFB 1313. We thank the the German Research Foundation (DFG) for supporting this work by funding EXC 2075/1-390740016 under Germany's Excellence Strategy as well as the Center for Digitalization and Technology Research of the Armed Forces of Germany (dtec.bw) through the project Macro/Micro-Simulation of Phase Decomposition in the Transcritical Regime (MaST); dtec.bw is funded by the European Union--NextGenerationEU\@. We acknowledge support by the Stuttgart Center for Simulation Science (SimTech).

\section*{Data availability statement}
\textcolor{red}{The data set is currently unpublished and it will be published with the final version of this manuscript:} 
The data that support the findings of this study are openly available in the data repository of the University of Stuttgart (DaRUS) at \textcolor{red}{[Doi/Url]}, reference number \textcolor{red}{[reference number]}.

\section*{\label{sec:appendix_entropyScaling} Appendix A: Entropy Scaling for Viscosity and Diffusion Coefficients}

Based on the (dimensionless) residual entropy profile, the viscosity and self-diffusion coefficient profiles can be determined using ansatz functions according to 
\begin{subequations}
\begin{equation} \label{eq:poly3}
  \ln \left(\frac{\eta^\mathrm{entr. scal.}}{\eta_\mathrm{ref}}\right)= \sum_{i}^{N} x_i A_i + \sum_{i}^{N} \frac{x_i m_i}{\tilde{m}} B_i s_\mathrm{res}^\#   + \sum_{i}^{N} \frac{x_i m_i}{\tilde{m}} C_i  \left(s_\mathrm{res}^\#\right)^2  + \sum_{i}^{N} \frac{x_i m_i}{\tilde{m}} D_i \left(s_\mathrm{res}^\#\right)^3
\end{equation}
\begin{equation}
  \ln \left(\frac{\rho D_i^{\text {self}, 0}}{\rho D_{i, \mathrm{ref}}^{\text {self},0}}\right)=  A_i+B_i s_\mathrm{res}^\#-C_i\left(1-\exp \left(s_\mathrm{res}^\#\right)\right)\left(s_\mathrm{res}^\#\right)^2 -D_i\left(s_\mathrm{res}^\#\right)^4  -E_i\left(s_\mathrm{res}^\#\right)^8
  \end{equation}
\end{subequations}
where the dependence on the position of the quantities $\eta^\mathrm{entr. scal.}(\rb)$, $D_i^{\text {self}, 0}(\rb)$, $x_i(\rb)$ and $s_\mathrm{res}^\#(\rb)$ is not shown explicitly for clarity. The ansatz functions are analogous to those in homogeneous systems \citep{loetgeringlin2018pure,hopp2018self,hopp2019erratum,stierle2021hydrodynamic}. Importantly, the parameters $A_i, B_i, C_i, D_i$, and $E_i$ were fitted to transport coefficients in homogeneous systems \citep{loetgeringlin2018pure,hopp2018self} and can be used directly in inhomogeneous systems. Thus, transport coefficients of inhomogeneous systems are not required as input to the generalized entropy scaling approach. 
The references for viscosity and the self-diffusion coefficient are the Chapman-Enskog properties $\eta_\mathrm{ref}=\eta_\mathrm{CE}$ and $D_{i,\mathrm{ref}}^{\text {self},0}=D_{i,\mathrm{CE}}^{\text {self},0}$, given by \citep{chapman1990mathematical,hirschfelder1967molecular}
  \begin{subequations}
        \begin{align}
    \eta_{\mathrm{CE}}=&\sum_{i=1}^{N_{\mathrm{c}}} \frac{x_i \eta_{\mathrm{CE}, i}}{\sum_j^{N_{\mathrm{c}}} x_j \phi_{i j}} \\
    \eta_{\mathrm{CE}, i}=&\frac{5}{16} \frac{\sqrt{\frac{\mw_i k_{\mathrm{B}} T}{N_{\mathrm{A}} \pi}}}{\sigma_i^2 \Omega_i^{(2,2)\#}} \\
    \phi_{i j}=&\frac{\left(1+\sqrt{\frac{\eta_{\mathrm{CE}, i}}{\eta_{\mathrm{CE}, j}} \sqrt[4]{\frac{\mw_j}{\mw_i}}}\right)^2}{\sqrt{8\left(1+\frac{\mw_i}{\mw_j}\right)}} \\
    \rho D_{i, \mathrm{CE}}^{\mathrm{self}, 0}=&\frac{3}{8} \frac{\sqrt{\frac{\mathrm{R} T}{\pi \mw_i}}}{\sigma_i^2 \Omega_i^{(1,1)\#}}
  \end{align}
\end{subequations}
where the dimensionless collision integrals, $\Omega^{(1,1)\#}$ and $\Omega^{(2,2)\#}$, are determined from an empirical correlation \citep{neufeld1972empirical}, $N_\mathrm{A}$ is the Avogadro number, and $R$ is the universal gas constant. 

The Maxwell-Stefan diffusion coefficients in the  mixture, $D_{ij}$, are estimated from the self-diffusion coefficients, $D_i^{\text {self}, 0}$.  
The relation between the pure-component self-diffusion coefficient and the self-diffusion coefficient at infinite dilution can be estimated as reported in unpublished work reported by \citet{stierle2021hydrodynamic} according to 
\begin{equation}
  D_i^{\mathrm{self},\infty} = D_j^{\mathrm{self},0} \left( \frac{d_j^\mathrm{eff}}{d_i^\mathrm{eff}} \right)^{2.23}
\end{equation}
where effective molecular diameters, $d_i^\mathrm{eff}$, are determined from liquid densities at the normal boiling point for each pure component $i$ as 
\begin{equation}
  d_i^\mathrm{eff} = \left( \rho_i^\mathrm{pure}(T_i^\mathrm{sat}, p\stst) \right)^{-\tfrac{0.71}{3}}
\end{equation}
with the saturation temperature $T_i^\mathrm{sat}(p\stst)$ and the pressure $p\stst=\SI{1.01325}{\bar}$ at the normal boiling point.
The self-diffusion coefficient in the mixture (or \emph{tracer} diffusion coefficient) is calculated according to \citet{liu2011predictive}
\begin{equation}
  \frac{1}{D_i^\mathrm{self}(\mathbf{x})} = \frac{x_i}{D_i^{\mathrm{self},0}} + \frac{(1 - x_i)}{D_i^{\mathrm{self},\infty}}
\end{equation}
The Maxwell-Stefan binary diffusion coefficient is determined based on the Darken equation \citep{darken1948diffusion,sridhar2010commentary} according to
\begin{equation}
  D_{ij}(\mathbf{x}) = \left( x_i D_j^\mathrm{self}(\mathbf{x}) + x_j D_i^\mathrm{self}(\mathbf{x}) \right) 
\end{equation}

\section*{References}

\bibliography{massTransfer}

\newpage
\section*{Supporting Information: "Mass Transfer Through Vapor-Liquid Interfaces From Hydrodynamic Density Functional Theory" }

\subsection*{Equilibrium Densities for Higher Temperature}

\begin{figure}[ht]
  \centering
  \includegraphics[width=0.6\columnwidth]{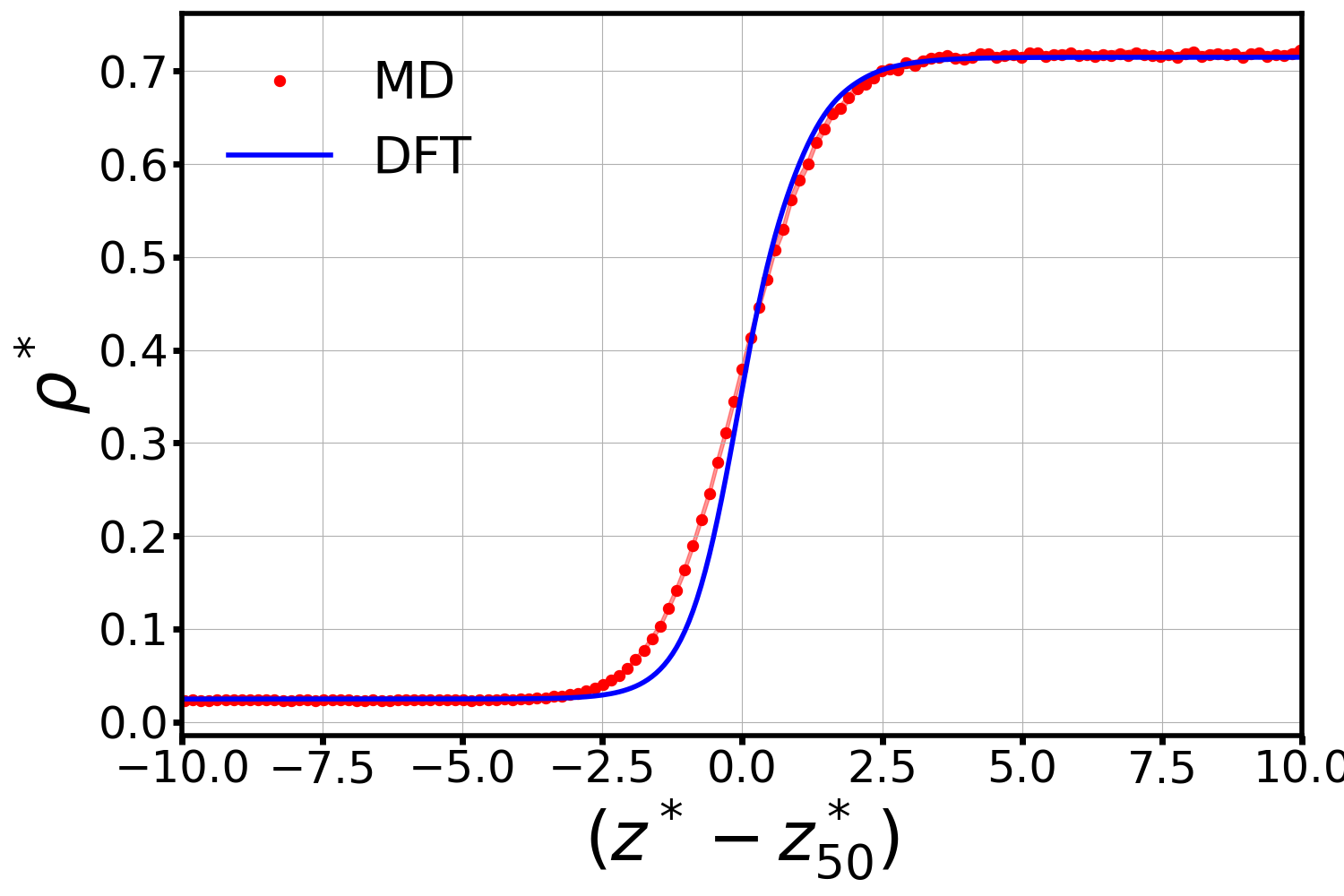}
  \caption{Equilibrium density profile of the pure system containing only component 1 from DFT (blue line) and MD (red symbols) at $T^*=0.825$. }
  \label{fig:rho_eq_pure}
\end{figure}

\begin{figure}
  \centering
  \subfloat[Mixture A.]{ \label{fig:rho_eq_mix_A}
      \centering
      \includegraphics[width=0.475\textwidth]{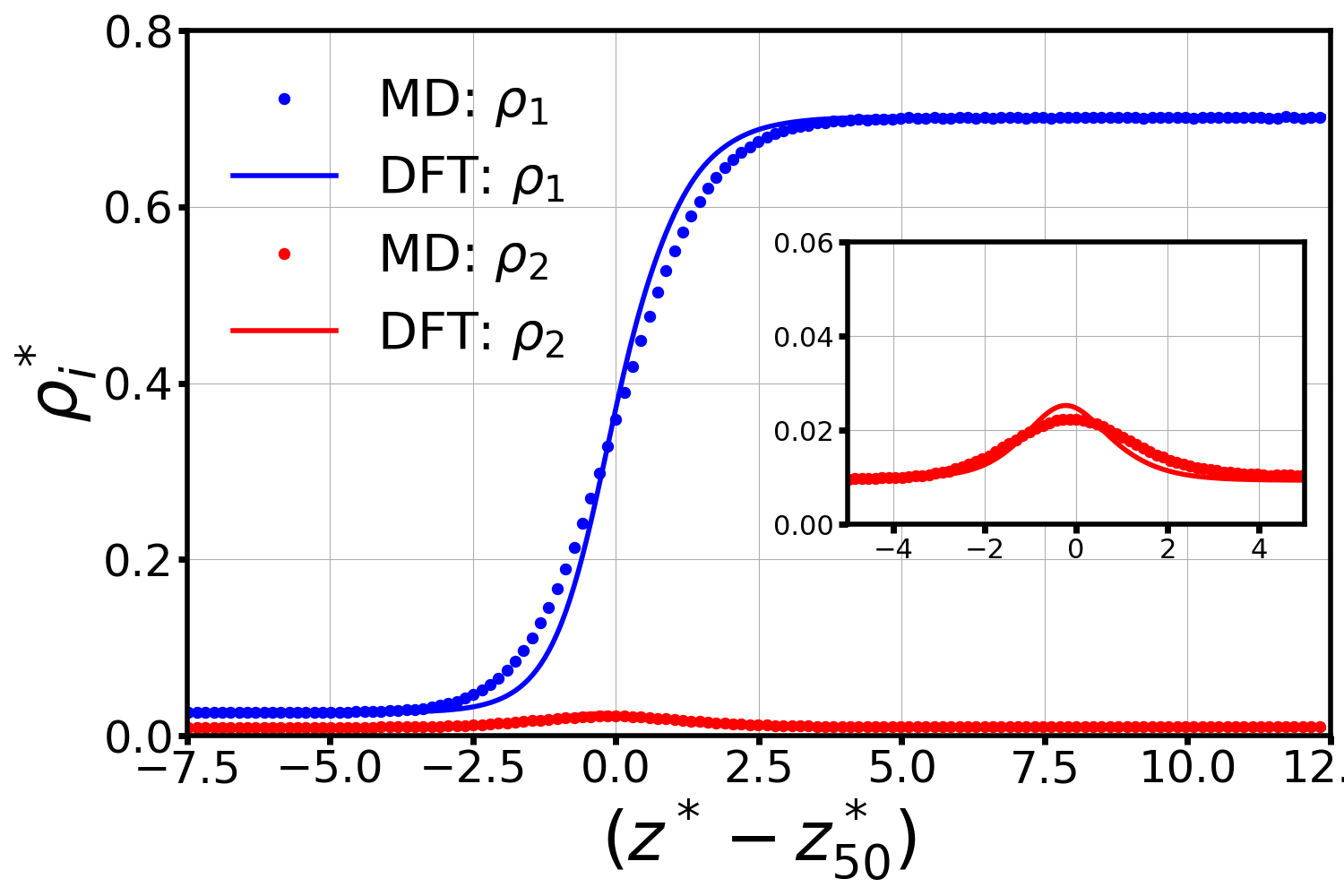}\label{fig:rho_static_dft}
 }
 \hfill
      \subfloat[Mixture B.]{  \label{fig:rho_eq_mix_B}
      \centering
      \includegraphics[width=0.475\textwidth]{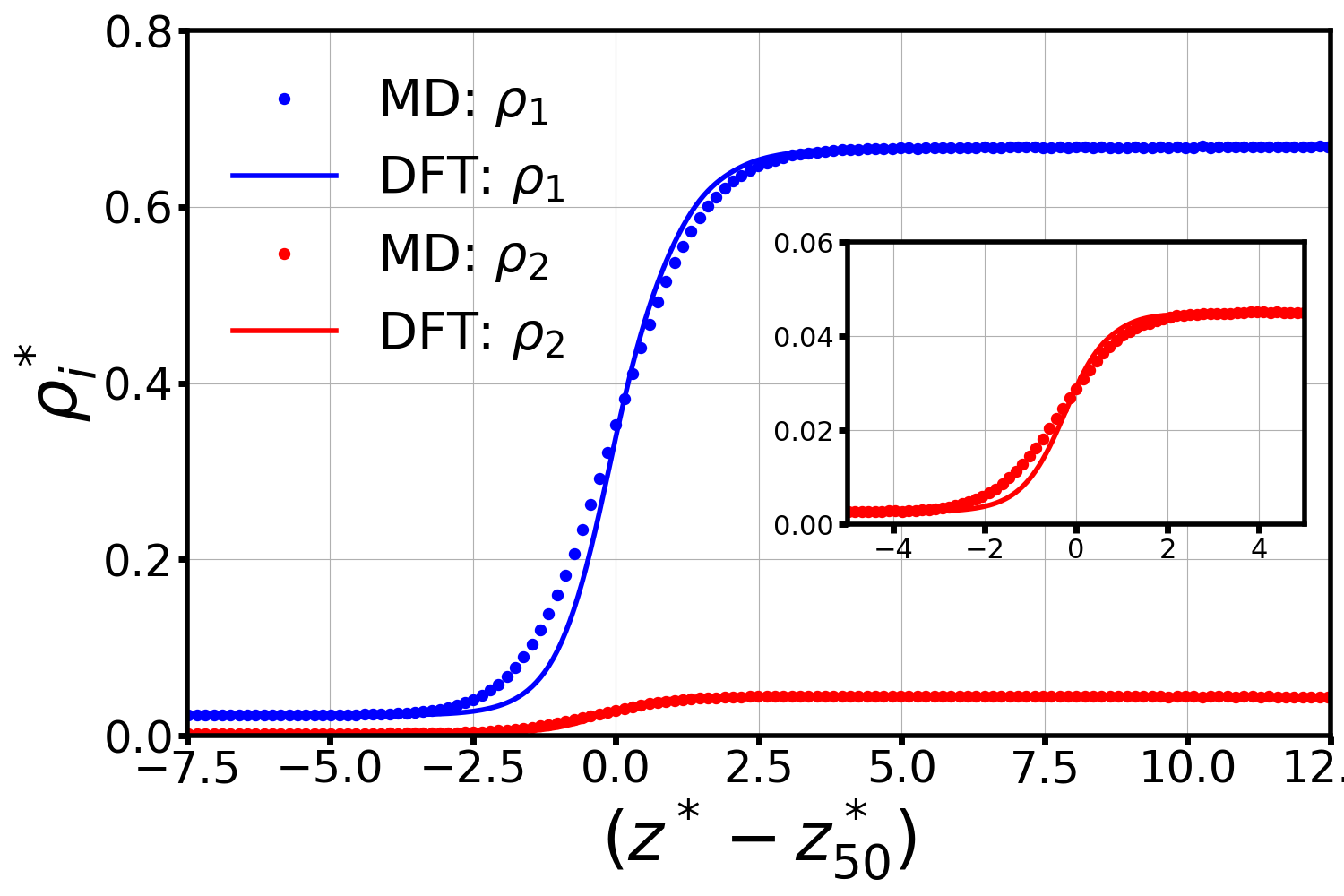}\label{fig:rho_static_emd}
 } 
 \caption{Equilibrium density profiles of both components in both mixtures from DFT (lines) and MD (points) at $T^*=0.825$. The inset magnifies the density of component 2. }
  \label{fig:rho_eq_mix}
\end{figure} 

This section provides equilibrium density profiles for $T^*=0.825$ for the pure system and both mixtures, determined from equilibrium DFT and MD simulations. \Cref{fig:rho_eq_pure} shows the equilibrium density for the pure system. Similarly to the lower temperature $T^*=0.715$, as provided in the main text, good agreement between DFT and MD is found. 
The density profiles for both components in each mixture are provided in \cref{fig:rho_eq_mix}. Again, similar results are observed compared to results for $T^*=0.715$ shown in the main text and good agreement is found between MD and DFT. 

\begin{figure}
  \centering
  \subfloat[Mixture A.]{ \label{fig:rho_eq_2_T_A}
      \centering
      \includegraphics[width=0.475\textwidth]{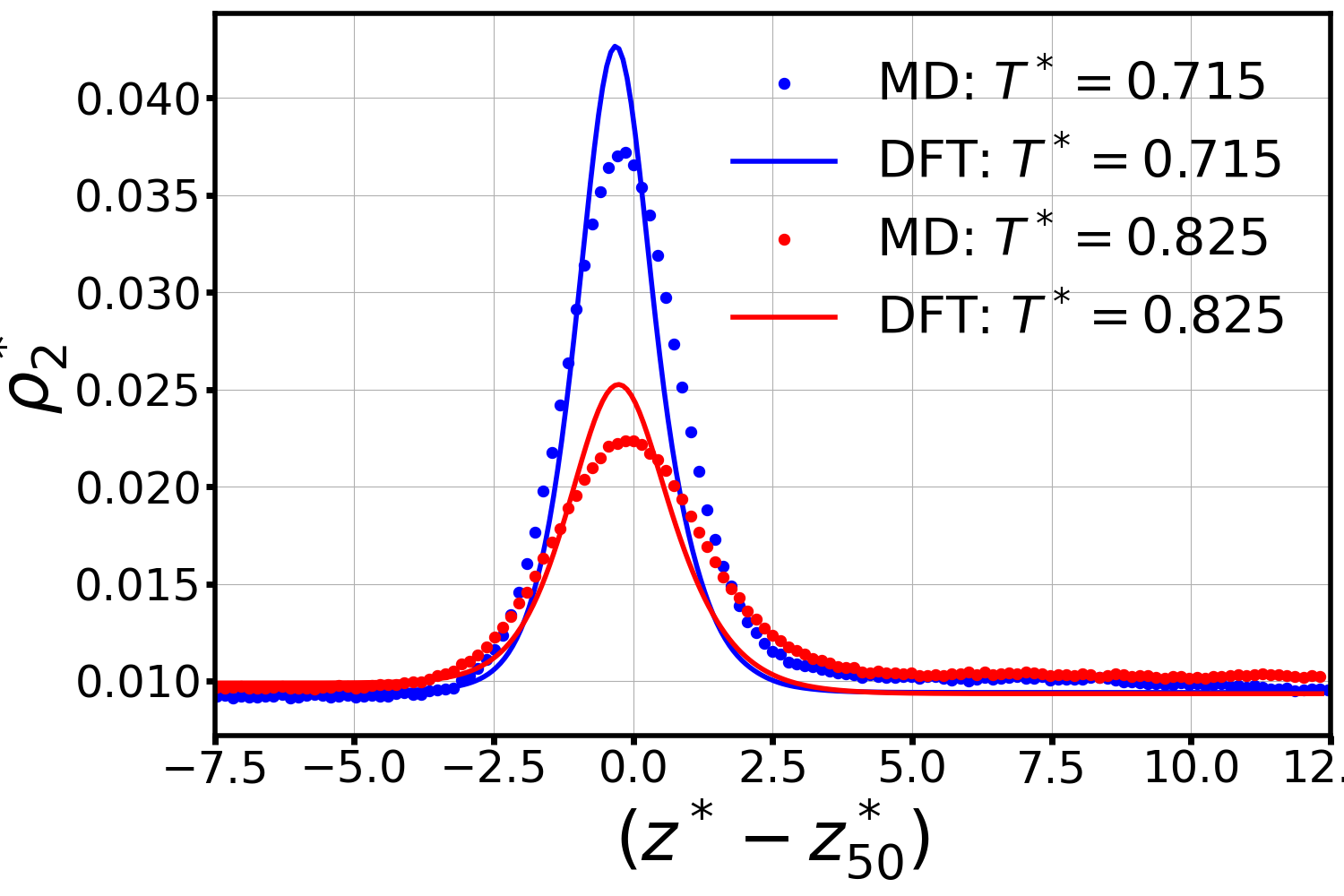}
 }
 \hfill
      \subfloat[Mixture B.]{  \label{fig:rho_eq_2_T_B}
      \centering
      \includegraphics[width=0.475\textwidth]{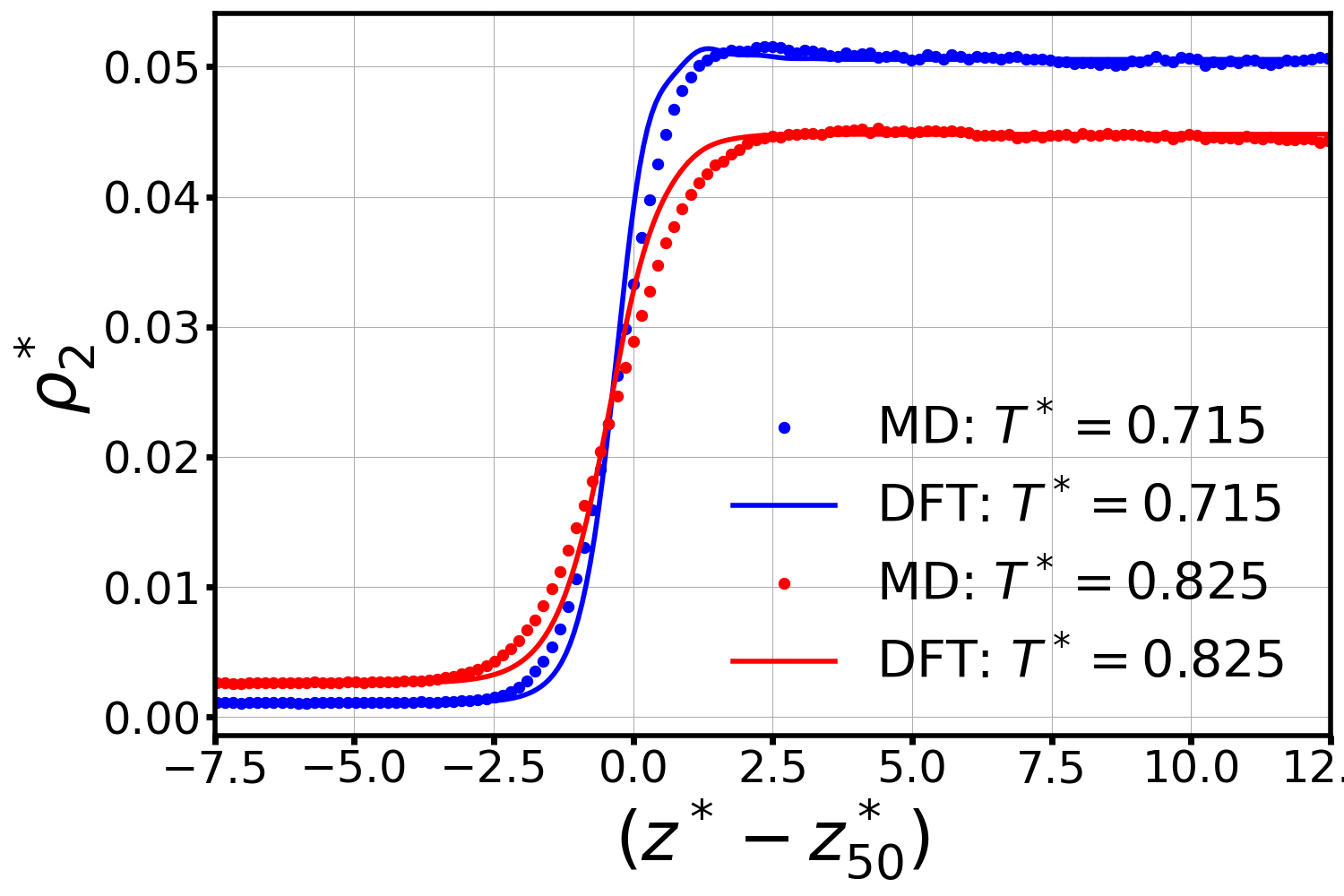}
 } 
 \caption{Equilibrium density of second component $\rho_2^*$  at different temperatures  from DFT (lines) and MD (points). }
  \label{fig:rho_eq_2_T}
\end{figure} 
\Cref{fig:rho_eq_2_T} compares the equilibrium density of the second component $\rho_2^*$ for the two different temperature $T^*=0.715$ and $T^*=0.825$ for both mixtures. The comparison clearly shows that the equilibrium densities in the bulk phases are almost independent of temperature for mixture A, while the opposite is true for mixture B. Furthermore, for the equilibrium densities in the liquid phase of mixture A  some deviations between DFT and MD are observed, which is due to deviations in the density from the PeTS equation of state  and  equilibrium MD. 

\subsection*{Fluxes for Different Temperatures}

\begin{figure*}
  \centering
  \begin{subfigure}[t]{0.475\textwidth}
    \centering
    \caption{Vapor: density $j_2^*$ in MV$_\text{vap}$ for mixture A.}
    \includegraphics[width=\textwidth,trim={0 0cm 0 0cm}, clip]{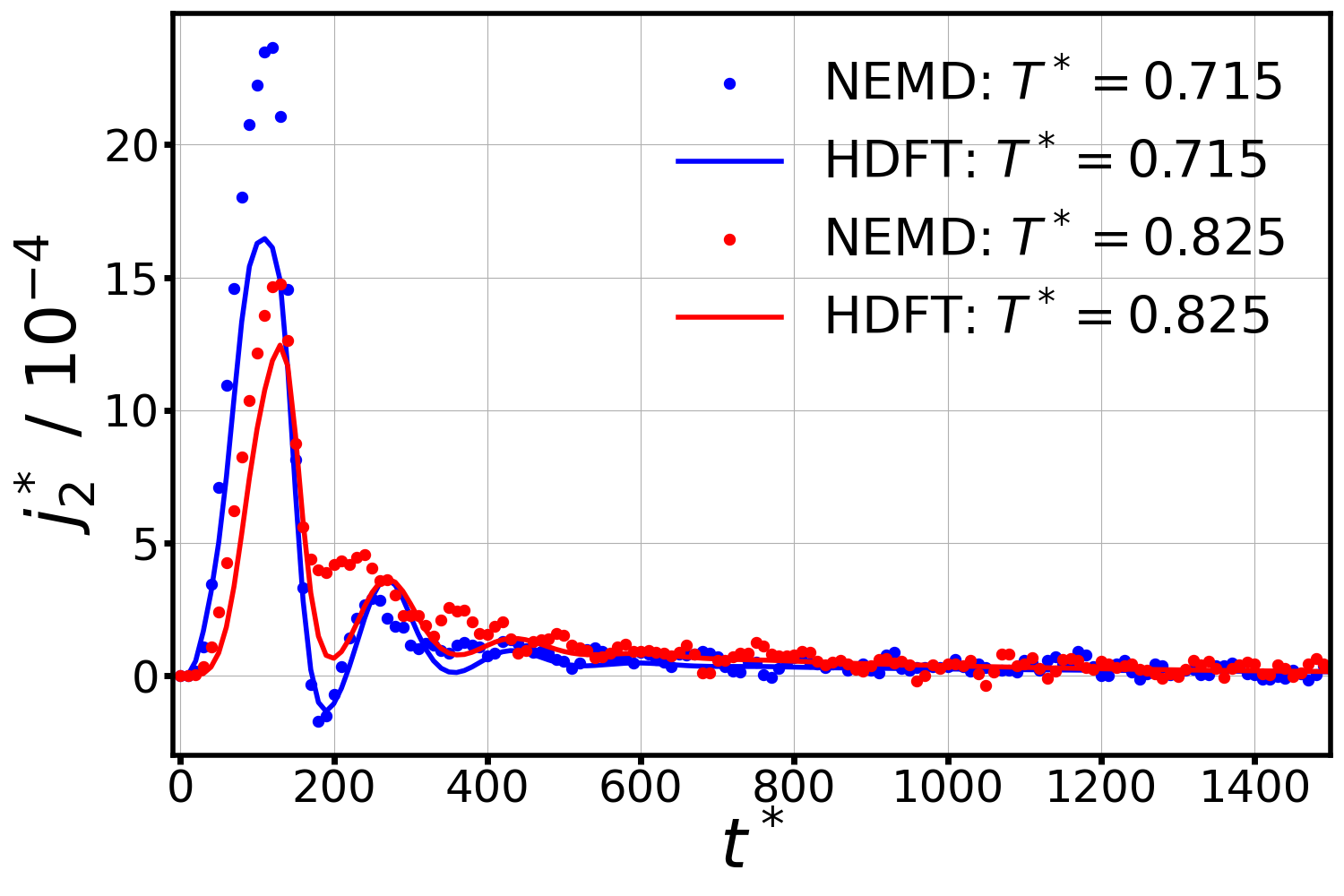}
    \label{fig:mv_T_rho_vap_mixA}
  \end{subfigure}
  \hfill
  \begin{subfigure}[t]{0.475\textwidth}
    \centering
    \caption{Liquid: density $j_2^*$ in MV$_\text{liq}$ for mixture A.}
    \includegraphics[width=\textwidth,trim={0 0cm 0 0cm}, clip]{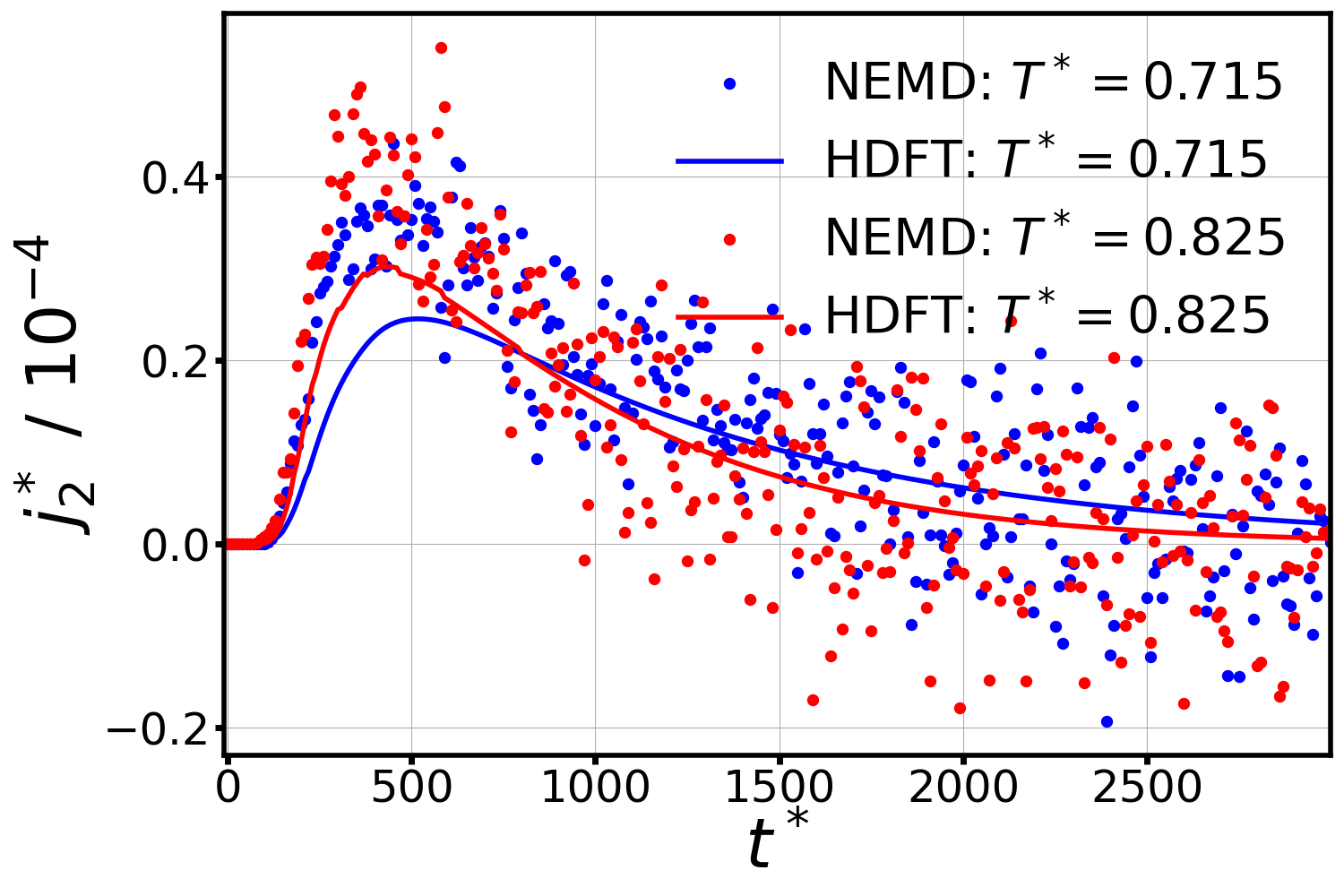}
    \label{fig:mv_T_rho_liq_mixA}
  \end{subfigure}
  \vfill
  \begin{subfigure}[t]{0.475\textwidth}
    \centering
    \caption{Vapor: density $j_2^*$ in MV$_\text{vap}$ for mixture B.}
    \includegraphics[width=\textwidth,trim={0 0cm 0 0cm}, clip]{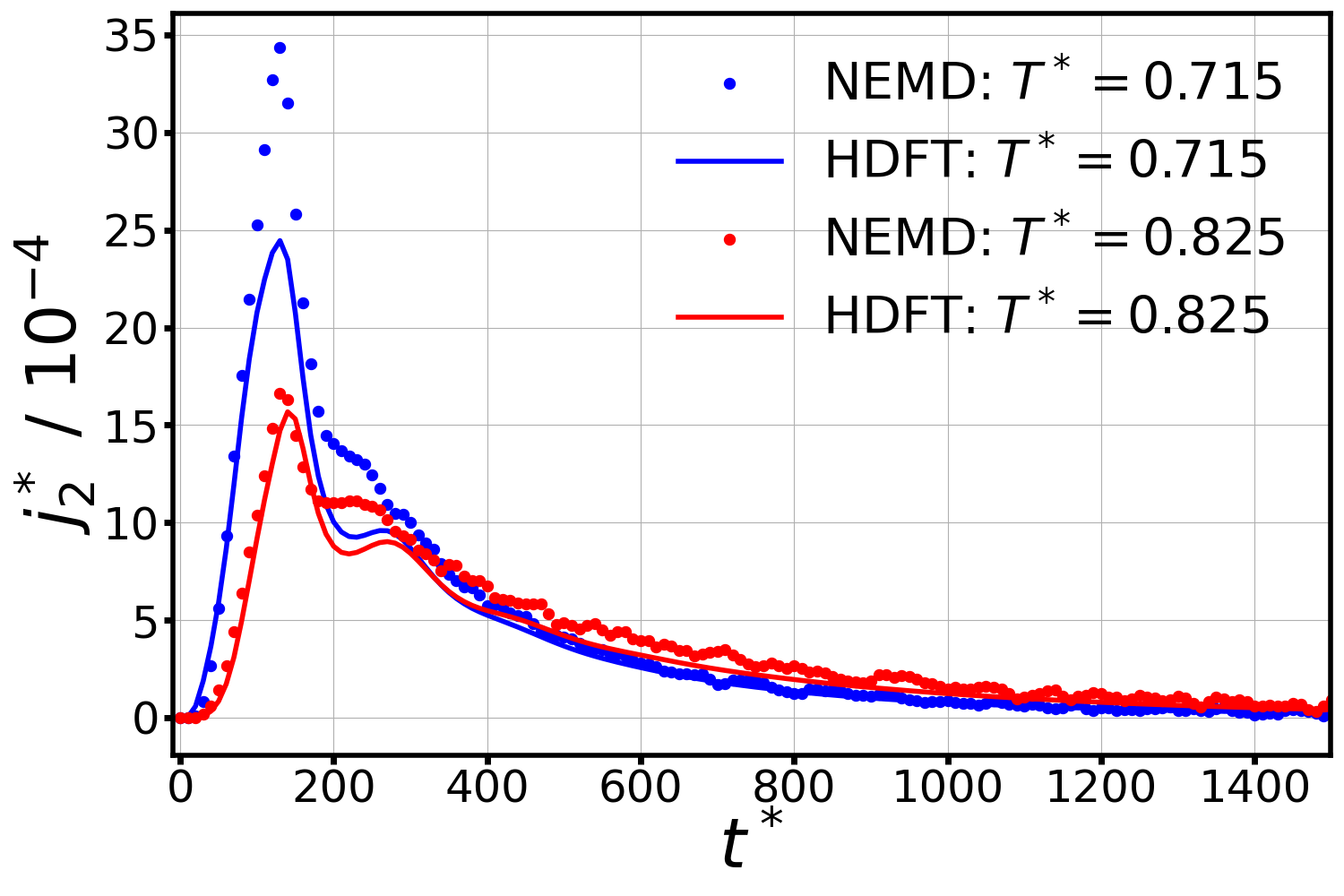}
    \label{fig:mv_T_rho_vap_mixB}
  \end{subfigure}
  \hfill
  \begin{subfigure}[t]{0.475\textwidth}
    \centering
    \caption{Liquid: density $j_2^*$ in MV$_\text{liq}$ for mixture A.}
    \includegraphics[width=\textwidth,trim={0 0cm 0 0cm}, clip]{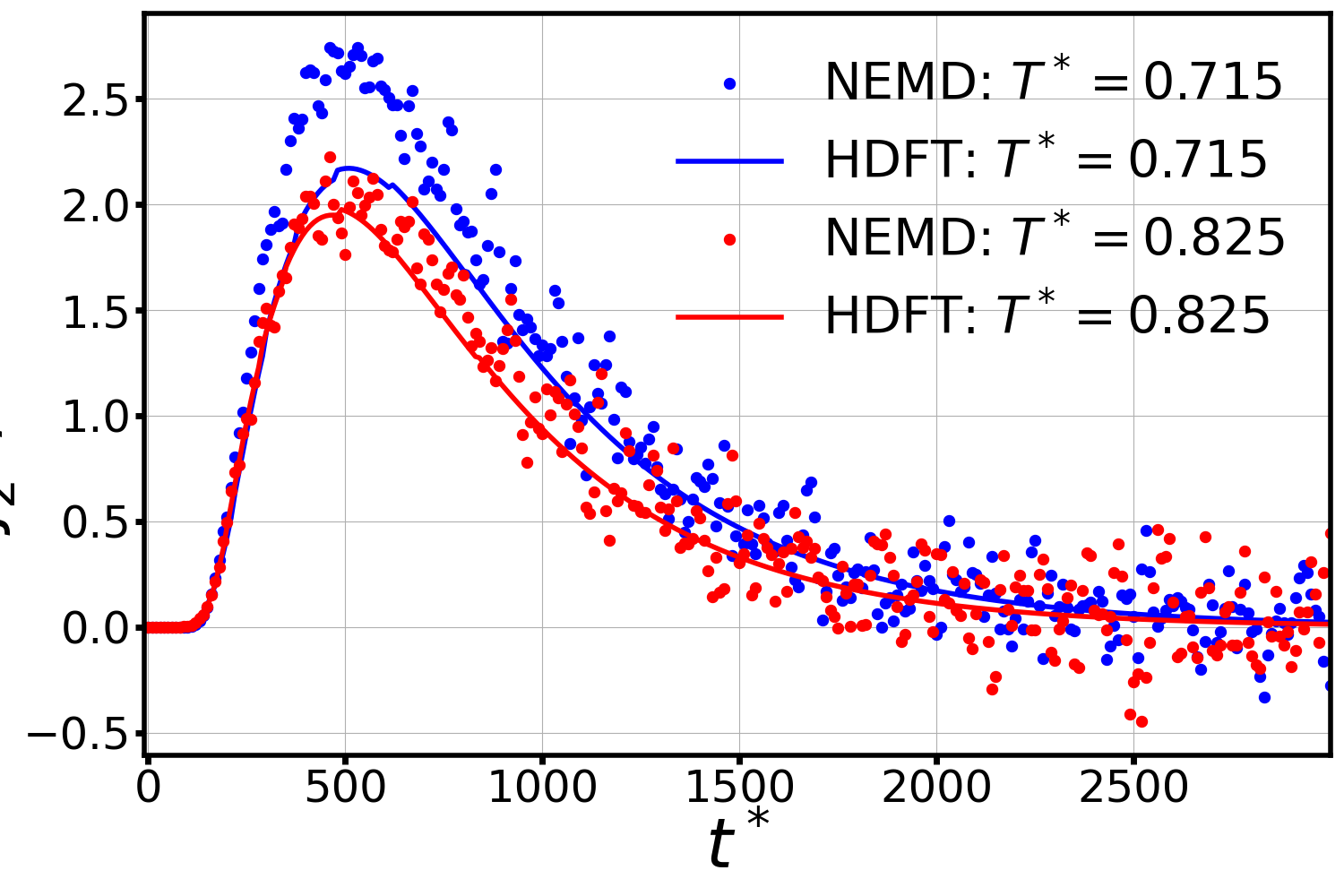}
    \label{fig:mv_T_rho_liq_mixB}
  \end{subfigure}
  \caption{Molecular flux $j_2^*$  of component 2 over time $t^*$ in measurement volumes MV$_\mathrm{vap}$ (left) and MV$_\mathrm{liq}$ (right) for  mixture A (top) and mixture B (bottom)  for $T^*=0.715$ (blue) and $T^*=0.825$ (red) from hydrodynamic DFT (lines) and NEMD (symbols).}
  \label{fig:mv_T_all}
\end{figure*}

The fluxes in the measurement volumes (MV) are shown in \cref{fig:mv_T_all}. Increasing the temperature reduces the height of the initial peak of the flux in MV$_\mathrm{vap}$. For mixture A (see \cref{fig:mv_T_rho_vap_mixA}) the flux oscillates and quickly approaches zero, whereas for mixture B (see \cref{fig:mv_T_rho_vap_mixB}) the flux remains positive for a longer time. In MV$_\mathrm{liq}$, the flux has a maximum and subsequently approaches zero for both mixture A and B, see \cref{fig:mv_T_rho_liq_mixA} and \cref{fig:mv_T_rho_liq_mixB}, respectively. However, the flux is slightly larger for the higher temperature in mixture A, while the opposite applies to mixture B. This agrees with the density profiles shown in the main text.
Results from hydrodynamic DFT show somewhat stronger  oscillations in MV$_\mathrm{vap}$ and slightly underestimate the maxima in MV$_\mathrm{liq}$. Overall, good agreement is observed between hydrodynamic DFT and NEMD for the fluxes and for both temperatures.

\subsection*{Flux and Density Profiles for Higher Temperature}

\begin{figure*}
  \centering
  \begin{subfigure}[t]{0.475\textwidth}
    \centering
    \caption{Mixture A: density $\rho_2^*$ profile.}
    \includegraphics[width=\textwidth,trim={0 0cm 0cm 0}, clip]{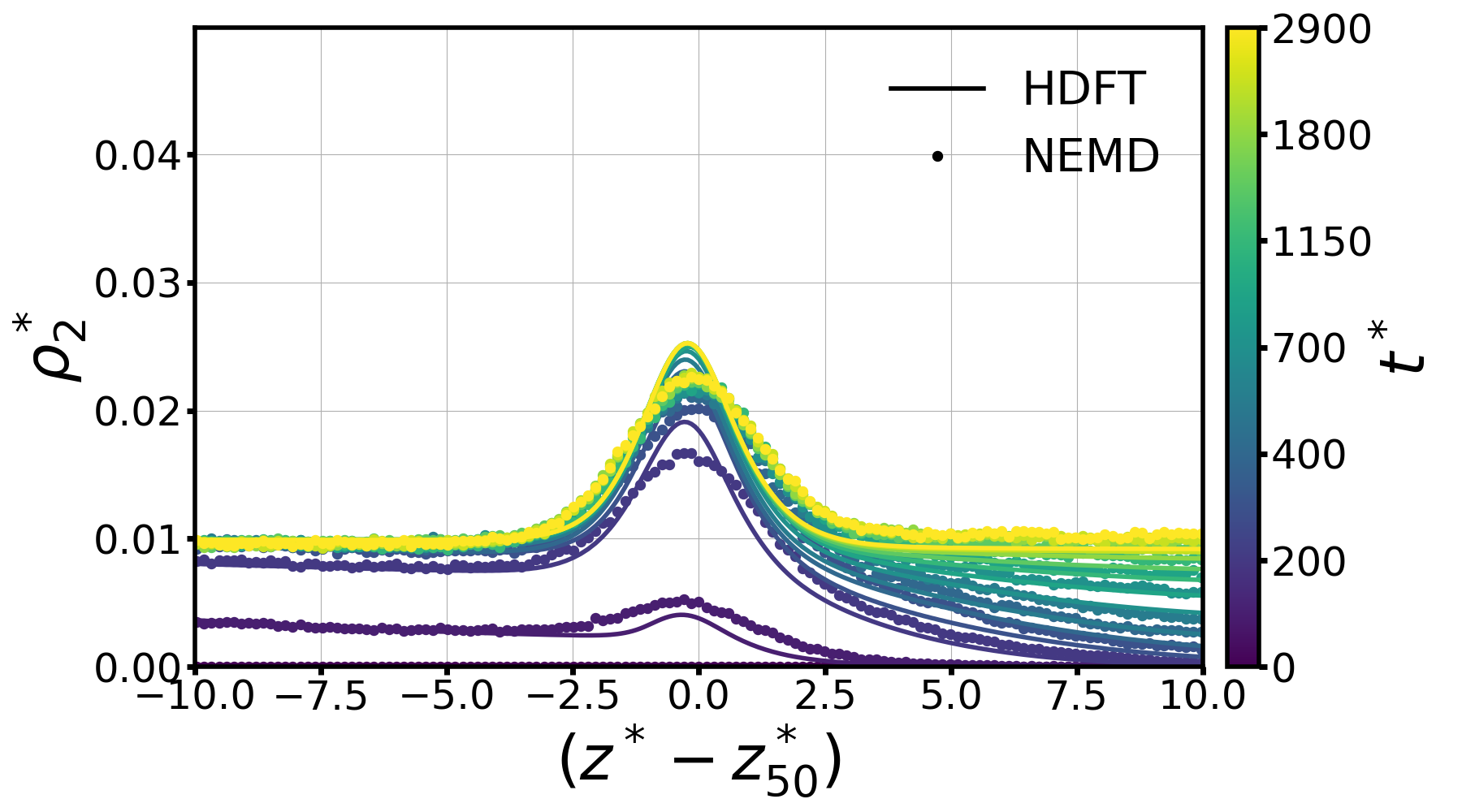}
    \label{fig:profiles_rho_A}
  \end{subfigure}
  \hfill
  \begin{subfigure}[t]{0.475\textwidth}
    \centering
    \caption{Mixture B: density $\rho_2^*$ profile.}
    \includegraphics[width=\textwidth]{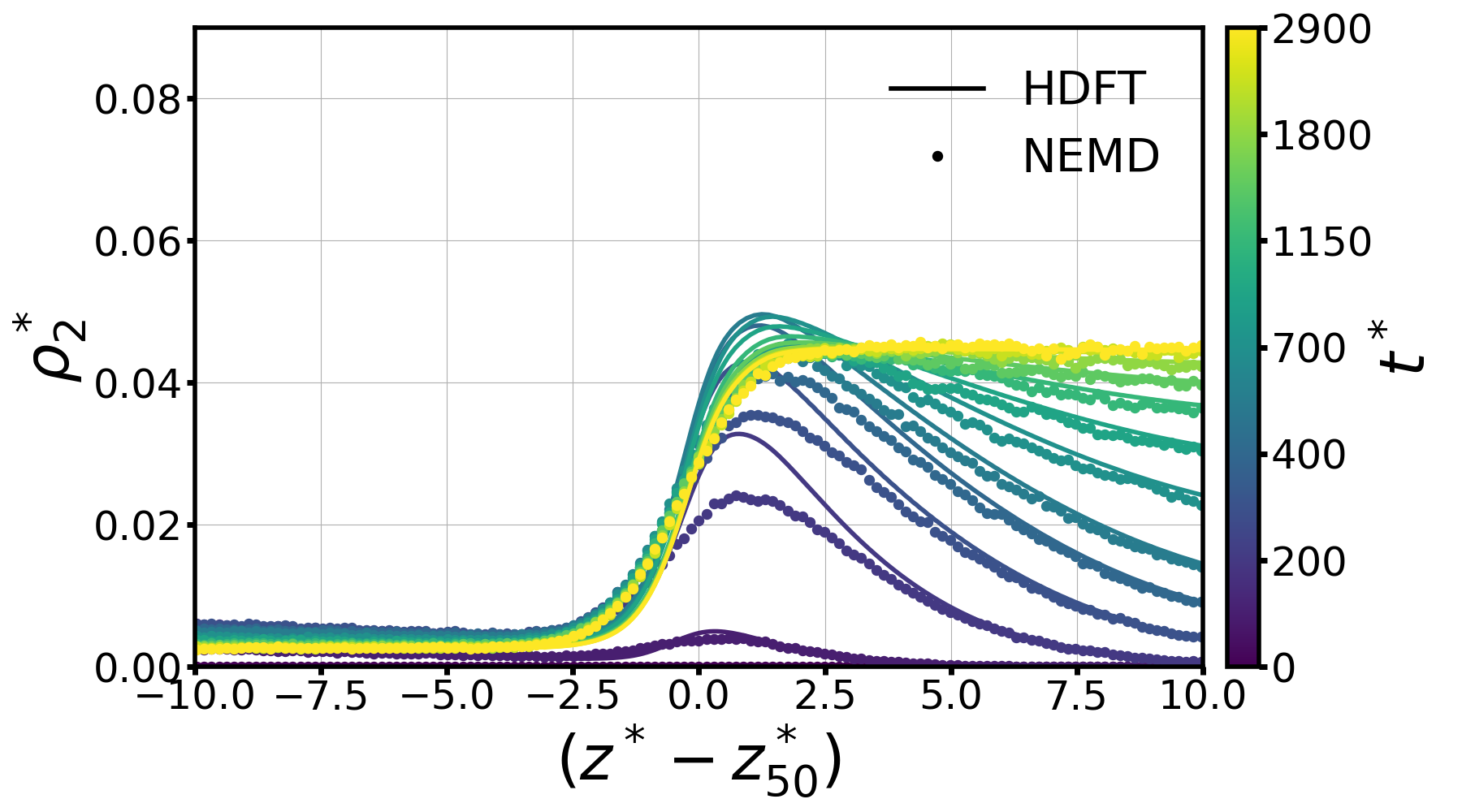}
    \label{fig:profiles_rho_B}
  \end{subfigure}
  \vfill
  \begin{subfigure}[t]{0.475\textwidth}
    \centering
    \caption{Mixture A: flux $j_2^*$ profile.}
    \includegraphics[width=\textwidth,trim={0 0cm 0 0cm}, clip]{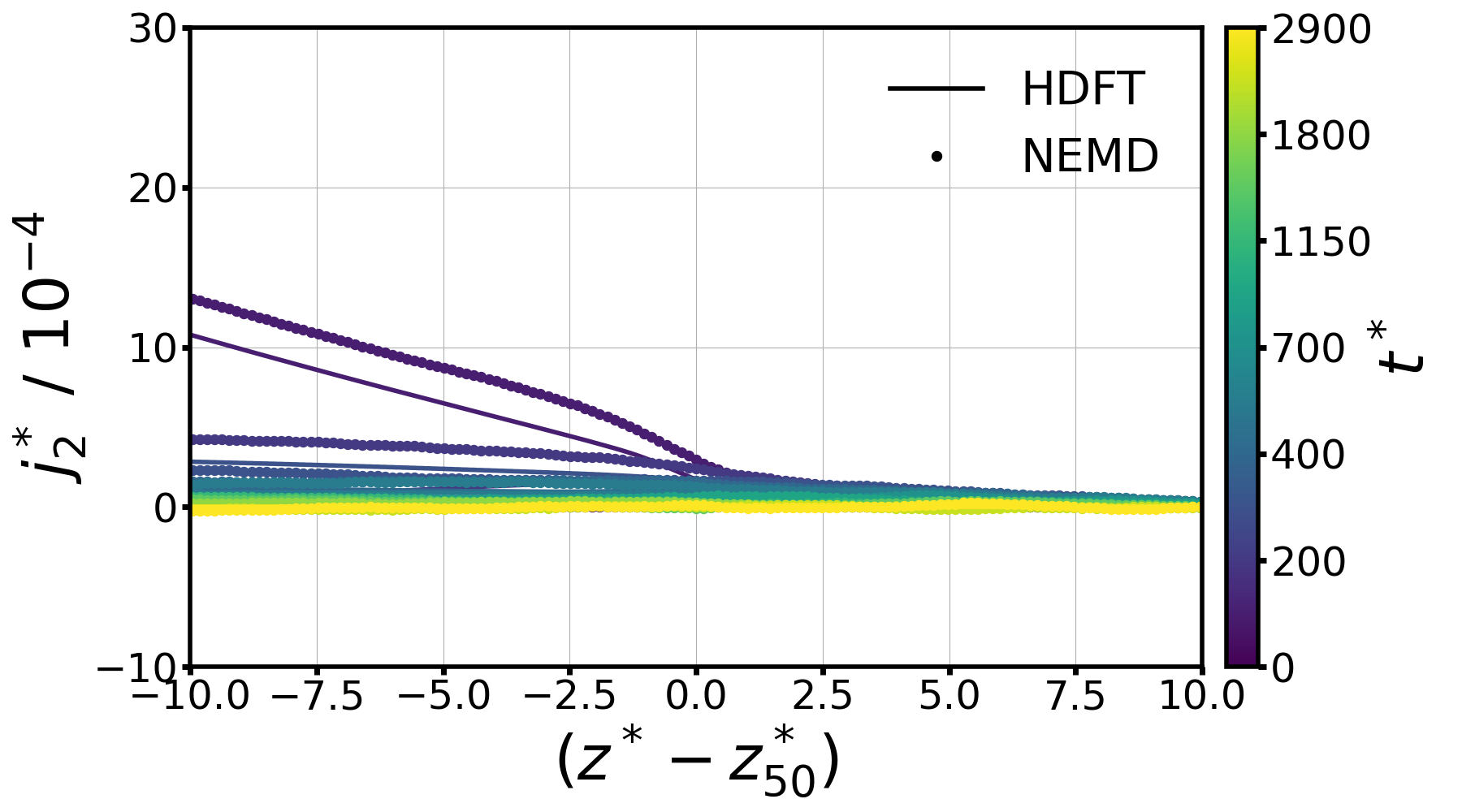}
    \label{fig:profiles_flux_A}
  \end{subfigure}
  \hfill
  \begin{subfigure}[t]{0.475\textwidth}
    \centering
    \caption{Mixture B: flux $j_2^*$ profile.}
    \includegraphics[width=\textwidth,trim={0 0cm 0 0cm}, clip]{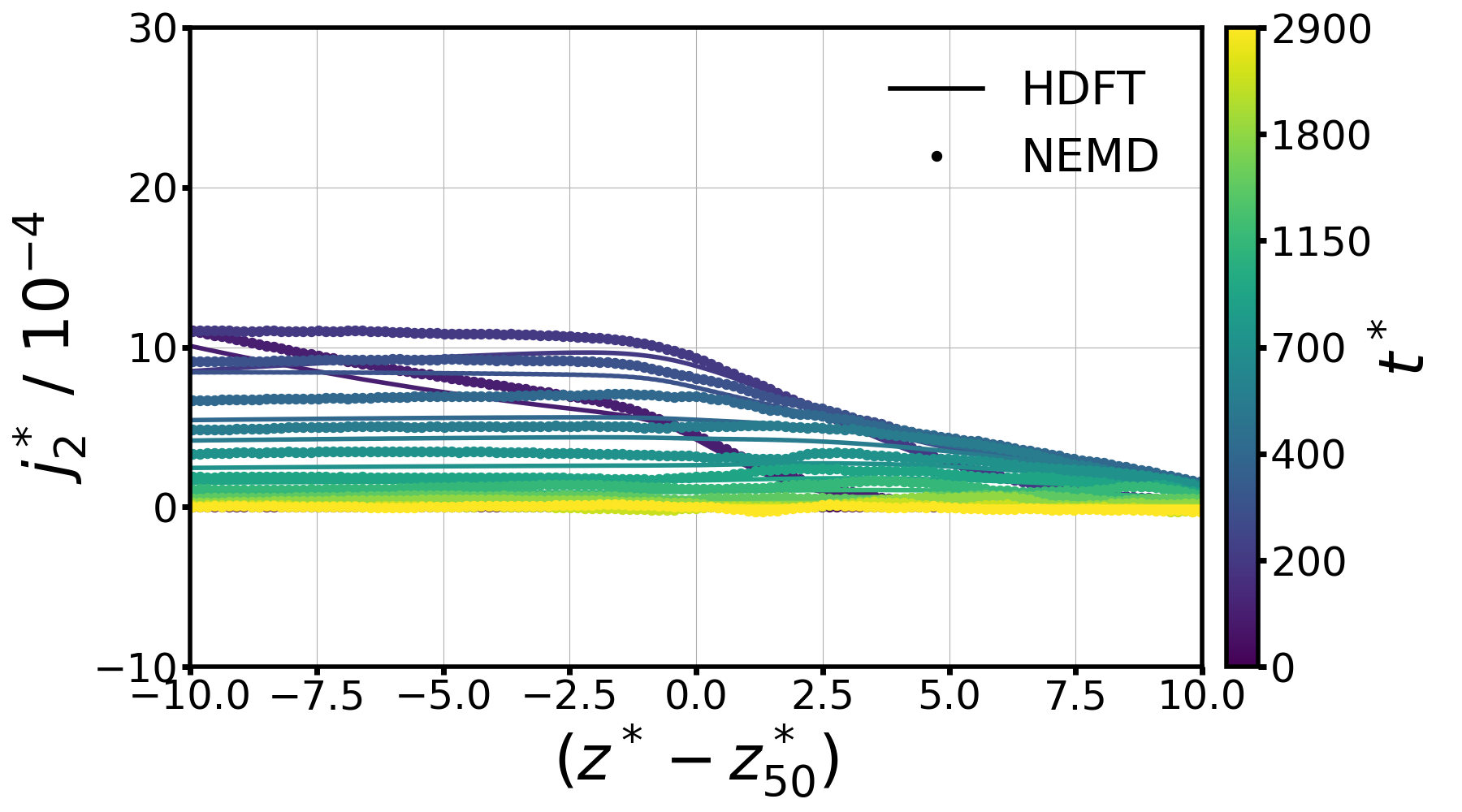}
    \label{fig:profiles_flux_B}
  \end{subfigure}
  \caption{Density $\rho_2^*$ and flux $j_2^*$ profiles across the interface during relaxation from hydrodynamic DFT (lines) and NEMD (points) at $T^*=0.825$.}
  \label{fig:profiles}
\end{figure*}

\Cref{fig:profiles} shows the density and flux profiles for $T^*=0.825$. Similarly to the results for $T^*=0.715$, as discussed in the main text, hydrodynamic DFT agrees well with NEMD results. The enrichment, both temporary for mixture B and permanent for mixture A, is less pronounced compared to the lower temperature.  

\end{document}